\documentclass[11pt,a4paper]{article}

\usepackage{url}
\usepackage[numbers]{natbib}
\usepackage{amsmath}
\usepackage{amssymb}
\usepackage{slashed}
\usepackage{color}
\usepackage[T1]{fontenc}
\usepackage{hyperref}
\usepackage[a4paper,margin=1in,footskip=0.25in]{geometry}
\usepackage{morefloats}
\usepackage{authblk}
\usepackage{graphicx}

\newcommand{\q}{\mathbf{q}}

\begin{document}


\title{Superallowed nuclear beta decays and precision tests\\ of the Standard Model}

\author[1, 2]{Mikhail Gorchtein}  
\author[3, 4]{Chien-Yeah Seng}
\affil[1]{Institut f\"ur Kernphysik, Johannes Gutenberg-Universit\"{a}t,\\
	 J.J. Becher-Weg 45, 55128 Mainz, Germany}
\affil[2]{PRISMA$^+$ Cluster of Excellence, Johannes Gutenberg-Universit\"{a}t,\\  Mainz, Germany\\  email: gorshtey@uni-mainz.de}
\affil[3]{Facility for Rare Isotope Beams, Michigan State University, East Lansing, MI 48824, USA}
\affil[4]{Department of Physics, University of Washington,
	Seattle, WA 98195-1560, USA\\  email: seng@frib.msu.edu}

\maketitle
\begin{abstract}
For many decades, the main source of information on the top-left corner element of the Cabibbo-Kobayashi-Maskawa quark mixing matrix $V_{ud}$ were superallowed nuclear beta decays with an impressive 0.01\% precision. This precision, apart from experimental data, relies on theoretical calculations in which nuclear structure-dependent effects and uncertainties play a prime role. This review is dedicated to a thorough reassessment of all ingredients that enter the  extraction of the value of $V_{ud}$ from experimental data. We tried to keep balance between historical retrospect and new developments, many of which occurred in just five past years. They have not yet been reviewed in a complete manner, not least because new results are a-coming. This review aims at filling this gap and offers an in-depth yet accessible summary of all recent developments. 
\end{abstract}


\tableofcontents

\section{Introduction}

Some discoveries are destined to change our understanding of Nature in a fundamental way. Yet, some of these discoveries have to wait until the scientific community is ready to accept them, at times for decades. Remarkably, in the case of nuclear decays this happened twice. The discovery of the spontaneous radioactivity of uranium salts by Becquerel in 1896 has brought him, Marie Curie-Sk\l{}odowska and Pierre Curie, who extended Becquerel's experiments, the 1903 Nobel prize. Yet, the same observation was made by Ni\'{e}pce de Saint-Victor almost 4 decades ahead of Becquerel - and of time. Studies of the newly discovered invisible rays originating from spontaneous decays of radioactive nuclei in the magnetic field led to the
first observation of $\beta$ rays by Rutherford and Soddy, and a general classification in terms of $\alpha$, $\beta$ and $\gamma$ rays. It was also understood that nuclear $\alpha$ and $\beta$ decays, apart from radioactive emission, were accompanied by a transmutation of chemical elements. 
This can arguably be seen as the beginning of physics of elementary particles. 

The second twist of fortune concerned beta decay and its most outstanding feature:  parity violation. Wu's experiment in 1957 \cite{Wu:1957my} confirmed the hypothesis of Lee and Yang \cite{Lee:1956qn} that parity may not be conserved in weak decays, bringing Lee and Yang the 1957 Nobel prize. It is less known that already back in 1928 Cox, McIlwraith and Kurrelmeyer observed an "Apparent Evidence of Polarization in a Beam of beta-Rays" \cite{Cox1928}. Apart from Chase who improved the experimental technique and confirmed their findings \cite{Chase1929,Chase1930-1,Chase1930-2}, this work has been completely ignored, as
such a possibility was not taken any seriously at that time. 


For years to come since its discovery, beta decays furnished us with many crucial ingredients of what is called nowadays the Standard (electroweak) Model. The continuous beta spectrum could be explained with the existence of the neutrinos. Assuming that they indeed existed, 
Fermi formulated his contact theory of beta decay \cite{Fermi:1934hr}. It was an extremely successful tool to describe the phenomenology of beta decays, explaining decay rates with just one constant $G_F$ and simple selection rules. 
Even though Fermi's formalism was reminiscent of that used to compute the hydrogen spectrum in QED, the interactions were clearly very different. The first attempt to unify the two while explaining the short range and weakness of the weak interaction by introducing a massive charged mediator dates back to Klein in 1938 \cite{Klein1938}. The mass of the mediator being directly related to the range of the respective interaction \cite{Yukawa:1935xg}, and identifying the weak coupling constant with the electromagnetic one, the mass of this hypothetical particle would naturally come out as $M_W\sim\sqrt{4\pi\alpha\sqrt2/G_F}\sim100$\,GeV. 

It took 3 decades more and the effort of the community and its best minds to formulate a consistent theory of electroweak interaction \cite{Glashow:1961tr,Weinberg:1967tq,Salam:1968rm} with the symmetry breaking mechanism \cite{Higgs:1964pj,Englert:1964et} to generate the heavy gauge boson masses. 
Joined with quantum chromodynamics (QCD), the theory of strong interaction \cite{Fritzsch:1973pi}, the electroweak theory makes part of the Standard Model (SM). 
SM is a non-abelian gauge field theory based on a simple symmetry group $SU(3)_\mathrm{c}\times SU(2)_L\times U(1)_Y$; empowered by renormalizability \cite{tHooft:1972tcz}, 
asymptotic freedom \cite{Gross:1973id,Gross:1973ju,Gross:1974cs,Politzer:1973fx}, and many further fundamental concepts, it permits calculating observables with any desired precision, at least in principle. 


Despite being extremely successful, SM falls short of many observed phenomena. SM is agnostic of the nature of
dark energy and dark matter, or the origin of the matter-antimatter asymmetry in the universe. Even in particle physics, neutrino masses and the finite mass of the Higgs boson remain unexplained. Experiments at the precision frontier play a crucial role in our quest for physics beyond the Standard Model (BSM): Observables are measured to an extremely high precision, with the hope to observe small deviations from SM predictions. Statistically significant deviations can be interpreted as signals of BSM physics. Physical processes that satisfy any one (or both) of the following conditions are preferred: (1) Forbidden or highly-suppressed in SM, or (2) Allowed, but the SM prediction is very precise. In this review we focus on tests of SM with experiments that belong to the second class. Precise beta decays provide stringent tests of universality of weak interaction, as well as the completeness of SM.


Universality of the weak interaction, the equality of the Fermi constant as felt by leptons and quarks, is one of the cornerstones of SM. It finds its exact mathematical expression in the unitarity of the Cabibbo-Kobayashi-Maskawa (CKM) matrix \cite{Cabibbo:1963yz,Kobayashi:1973fv},
\begin{equation}
\left(\begin{array}{ccc}
V_{ud} & V_{us} & V_{ub}\\
V_{cd} & V_{cs} & V_{cb}\\
V_{td} & V_{ts} & V_{tb}
\end{array}\right)\left(\begin{array}{ccc}
V_{ud}^{*} & V_{cd}^{*} & V_{td}^{*}\\
V_{us}^{*} & V_{cs}^{*} & V_{ts}^{*}\\
V_{ub}^{*} & V_{cb}^{*} & V_{tb}^{*}
\end{array}\right)=1~,
\end{equation} 
which implies a number of constraints the CKM matrix elements must obey. 
The top-row unitarity constraint, $|V_{ud}|^2+|V_{us}|^2+|V_{ub}|^2=1$, is currently tested at 0.01\% level. Using the same rationale that led Klein to the expectation $M_W\sim100$\,GeV, such precision translates into probing BSM physics at very high scale,
\begin{equation}
\left(\frac{v_\text{H}}{\Lambda_\text{BSM}}\right)^2\lesssim 0.01\% \implies \Lambda_\text{BSM}\gtrsim 20~\text{TeV}~,
\end{equation}
comparable to those accessible at the Large Hadron Collider (LHC). 


To perform a unitarity test, one needs to study the appropriate charged weak (CW) decay processes that involve $V_{ud}$ and $V_{us}$ ($|V_{ub}|^2\sim 10^{-5}$ can safely be dropped given the precision goal). The primary avenues for $V_{ud}$ extraction are beta decays of the pion, free neutron and atomic nuclei, whereas $V_{us}$ can be obtained from leptonic or semileptonic decays of kaon, hadronic decay modes of the tau lepton, or from hyperon decays. To extract the CKM matrix elements, the respective decay rates, and possibly other spin-dependent quantities, must be measured with high precision. But that is not enough; we must compute all the SM theory corrections to the decay processes, at $10^{-4}$ level in the $V_{ud}$ sector, and $10^{-3}$ level in the $V_{us}$ sector, to ensure the accuracy and precision level of the extracted matrix elements. This is highly non-trivial as it requires a careful handle of hadronic and nuclear effects at the infrared scale where a perturbative solution of QCD is not available.

A combination of inputs from lattice QCD, nuclear many-body calculations and experiments are needed to properly pin down the relevant physics at low scale. These inputs have to be linked to the actual decay processes through appropriate theory prescriptions such as current algebra (CA), dispersion relation (DR), and effective field theory (EFT). In the recent years, we observe a tremendous improvement along this direction, in particular in the single-hadron sector. The first dispersive analysis of the single-nucleon $\gamma W$-box diagram~\cite{Seng:2018yzq} unveiled large systematic uncertainties of previous calculations. A more careful analysis of these uncertainties allowed to reduce them significantly; at the same time, it led to a substantially reduced central value of $V_{ud}$, an observation that was supported by several follow-up studies~\cite{Czarnecki:2019mwq,Shiells:2020fqp,Seng:2020wjq,Hayen:2020cxh}. The pioneering lattice QCD calculations of the mesonic $\gamma W$-box diagrams~\cite{Feng:2020zdc,Ma:2021azh,Yoo:2023gln} paved the way for similar calculations in the nucleon sector \cite{Ma:2023kfr}. A new analysis of $K_{\ell 3}$ radiative correction, empowered by the aforementioned lattice results, was able to reduce the theory uncertainty by an order of magnitude~\cite{Seng:2021boy,Seng:2021wcf,Seng:2022wcw}.

Significant progress is observed in the nuclear sector, as well. However, in contrast to the nucleon- and meson-level improvements that \textit{decrease} the theory uncertainties, new nuclear structure-dependent effects which were not accounted for in the past, \textit{increased} theory uncertainties in the nuclear beta decay. The dispersive analysis applied to nuclear decays unveiled important missing nuclear structure effects in radiative corrections~\cite{Seng:2018qru,Gorchtein:2018fxl}. Nuclear structure corrections at tree level, such as the isospin-symmetry-breaking (ISB) correction to the Fermi matrix element~\cite{Seng:2022epj,Seng:2023cvt} and subleading nuclear-structure dependent effects in Coulomb corrections~\cite{Seng:2022inj,Seng:2023cgl} have been re-examined, and the corresponding uncertainties are now again under careful scrutiny. For nearly 5 decades, the superallowed beta decays involving only $(J^P)=(0^+)$ nuclei have been the best avenue for extracting $V_{ud}$~\cite{Hardy:1975eq,Hardy:2020qwl}. Aforementioned developments in theory, joined with improved experimental precision in the neutron decay~\cite{Markisch:2018ndu,UCNt:2021pcg}, have changed the landscape, causing the free neutron decay to gradually catch up in the precision of $V_{ud}$ extraction. 
%
%
This puts more emphasis on a better understanding of nuclear-structure effects in superallowed beta decays where recent theory developments led to an increased uncertainty. Much effort has already been invested in devising ways to improve calculations and reduce their uncertainties through a combination of new experimental measurements and modern nuclear many-body calculations. This review  summarizes these new ideas and recent results.

\section{Basics of superallowed nuclear beta decays}

A beta decay refers to the following process, $\phi_i\rightarrow \phi_f\beta\nu$, where $\phi_{i,f}$ are strongly-bounded systems, i.e. hadrons or nuclei; $\beta$ here can be either $e^+$ ($\beta^+$-decay) or $e^-$ ($\beta^-$-decay). Restricting ourselves to nuclear systems, the ``zeroth-order'' interpretation of the $\beta^+$($\beta^-$) decay process is simply that a bounded proton (neutron) transitions weakly into a neutron (proton), and emits a lepton-neutrino pair. In the non-relativistic formalism, there are only two available single-nucleon transition operators when recoil corrections are neglected; they are proportional to $\tau_\pm$ and $\vec{\sigma}\tau_\pm$ respectively, where $\tau_\pm$ are the isospin rasing/lowering operator that changes the nucleon flavor, and $\vec{\sigma}$ are the Pauli spin matrices. Transitions that can be triggered by these two operators are known as ``allowed'' beta decays, and must satisfy the selection rules: $\Delta J=0,1$ with no parity change.  


The $0^+\rightarrow 0^+$ transition bears the name of ``superallowed'' beta decay and plays a special role in precision physics, both in the extraction of $V_{ud}$ and search of new physics. It is a pure Fermi transition, and the only relevant nuclear matrix element at tree level is the Fermi matrix element $\langle \phi_f|\tau_+|\phi_i\rangle$. In the limit of isospin symmetry, this matrix element is completely fixed by group theory and is nucleus-independent; this is also known as the conserved vector current (CVC) hypothesis.
The fact that the tree-level matrix element is not renormalized by QCD corrections allows for a high-precision extraction of $V_{ud}$. 


Entering the era of precision, the ability to compute the SM contributions to the beta decay rate with a $10^{-4}$ accuracy is mandatory in order to meet the increasing experimental precision and to maximize the sensitivity to BSM physics. 
The master formula that serves as the basis for the $V_{ud}$ extraction and tests of SM with superallowed beta decays was introduced by Hardy and Towner~\cite{Hardy:1975eq} and  reads:
\begin{equation}
\mathcal{F}t=\frac{K}{2V_{ud}^2G_F^2(1+\Delta_R^V)}~.\label{eq:master}
\end{equation}
This formula is carefully structured so that all the nucleus-dependent quantities (theory and experiment) are combined on the left-hand side, while only the nucleus-independent ones are on the right-hand side.
Apart from $V_{ud}$, it contains 
the factor $K/(\hbar c)^6=2\pi^3\hbar \ln 2/(m_e c^2)^5=8120.27648(26)\times 10^{-10}~\text{GeV}^{-4}\text{s}$ and the Fermi constant $G_F/(\hbar c)^3=1.1663788(6)\times 10^{-5}~\text{GeV}^{-2}$, both very precisely measured~\cite{ParticleDataGroup:2022pth}. The remaining ingredient is $\Delta_R^V$ \cite{Wilkinson:1970cdv}, the universal radiative correction (RC) to the vector coupling, which was recently reviewed in detail in \cite{Gorchtein:2023srs}, and is briefly discussed in Subsection \ref{sec:Boxn}.

To be equalled to the nucleus-independent right-hand side of Eq.\eqref{eq:master}, the combination on its left-hand side must be nucleus-independent, as well. 
All nucleus-dependent quantities are absorbed into the ``modified $ft$-value" 
denoted by $\mathcal{F}t$, 
\begin{equation}
\mathcal{F}t=ft(1+\delta_\text{R}')(1+\delta_\text{NS}-\delta_\text{C})~.\label{eq:Ft}
\end{equation} 
Its universality (i.e., nucleus-independence) is a consequence of CVC and of absence of non-vector  BSM interactions. The entries in Eq.~\eqref{eq:Ft} reflect the history of the beta-decay theory.
The statistical rate function $f$ dates back to Fermi's 1934 article \cite{Fermi:1934hr}, and is numerically the largest. The pioneering Fermi's calculation has been improved upon over the following decades by many authors, and we refer the reader to a recent comprehensive review \cite{Hayen:2017pwg}.
The remaining terms are generally small, $\ll1$. The outer RC 
 $\delta_\text{R}'$ at leading order in $\alpha$ is the universal Sirlin's function \cite{Sirlin:1967zza}, but was found by Jaus and Rasche to retain a significant dependence on the daughter-nucleus charge $Z$ at higher orders \cite{Jaus:1970tah,Jaus:1972hua}.
 However, it is calculable in QED and does not depend on the nuclear structure. 
The Coulomb or, more generally, isospin-symmetry breaking (ISB) correction $\delta_\text{C}$ to the tree-level Fermi matrix element was introduced by MacDonald in 1958 \cite{MacDonald:1958zz}. Finally, the nuclear-structure correction was 
introduced by Jaus and Rasche in 1989
\cite{Jaus:1989dh}. The ingredients of $\mathcal{F}t$-values that depend on the nuclear structure are the main focus of this review.
%
For notational simplicity, we will adopt the natural units, namely $c=\hbar=1$, while keeping the electron mass $m_e$ explicit to better connect with relativistic notations.

\section{Tree-level nuclear structure effects}



In a superallowed nuclear beta decay, a proton in the nucleus transitions weakly into a neutron by emitting a positron-neutrino pair mediated by a $W$-boson. This transition is triggered by the hadronic CW current $J_W^{\dagger\mu}$; below we provides its relativistic expression in terms of the quark operators, and compare it to the electromagnetic (EM) current $J_\text{em}^\mu$, 
\begin{equation}
J_W^{\dagger\mu}=\bar{d}\gamma^{\mu}(1-\gamma_5)u~,~J_\text{em}^\mu=\frac{2}{3}\bar{u}\gamma^\mu u-\frac{1}{3}\bar{d}\gamma^\mu d-\frac{1}{3}\bar{s}\gamma^\mu s~.\label{eq:currents}
\end{equation}
The matrix element of $J_W^{\dagger 0}$ with respect to static nuclear external states is parametrized as
\begin{equation}
\langle \phi_f|J_W^{\dagger 0}(\vec{x})|\phi_i\rangle\equiv M_F \rho_\text{cw}(\vec{x})~,\label{eq:treeME}
\end{equation}
where $\rho_\text{cw}(\vec{x})$ is the CW distribution function (normalized to unity), and $M_F$ is the Fermi matrix element which can also be defined as $M_F\equiv \langle \phi_f|\tau_+|\phi_i\rangle$~, with 
\begin{equation}
\tau_+=\int d^3x (J_W^{\dagger 0}(\vec{x}))_V
\end{equation}
the isospin-raising operator. The remainder of this section will be devoted to the discussion of the nuclear effects that enter the distribution $\rho_\text{cw}(\vec{x})$ and the Fermi matrix element $M_F$.

\subsection{Statistical rate function}

We start by discussing the statistical rate function $f$ in Eq.\eqref{eq:Ft} It is usually introduced to encode all the nuclear and atomic structure effects at tree level in the isospin limit. In a generic allowed beta decay, there are about 12 such corrections (see, e.g. Ref.\cite{Hayen:2017pwg}), but for superallowed beta decays only a small subset of them is relevant. Following Hardy and Towner~\cite{Hardy:2004id,Hardy:2008gy}, we parameterize $f$ as:
\begin{equation}
f=m_e^{-5}\int_{m_e}^{E_0}|\vec{p}_e|E_e(E_0-E_e)^2F(E_e)C(E_e)Q(E_e)R(E_e)r(E_e)dE_e~,\label{eq:fformula}
\end{equation}
with $E_0\equiv M_i-M_f$ the leading-order electron end-point energy. Here we arrange the nuclear/atomic structure-dependent functions with decreasing degree of importance: The Fermi function $F(E_e)$, the shape factor $C(E_e)$, the atomic electron screening factor $Q(E_e)$, the kinematic recoil factor $R(E_e)$, the atomic overlap correction factor $r(E_e)$. 

\subsubsection{Hidden nuclear structure uncertainties}

Among the corrections above, only the Fermi function $F(E_e)$ and the shape factor $C(E_e)$ are sensitive to the details of the nuclear structure. The former characterizes the outgoing positron moving in the Coulomb field of the daughter nucleus~\cite{Fermi:1934hr}. Meanwhile, the latter encodes the spatial distribution of the nucleons that undergo the beta transition inside the nucleus.
Despite their close analogy, the two quantities received significantly different treatment in the literature.  The nuclear charge distribution of the daughter nucleus is everything one needs to know to compute the Fermi function in terms of the radial Coulomb functions 
obtained by solving the Dirac equation for the positron in the Coulomb field of the daughter nucleus~\cite{BehrensBuhring}.
The mean square (ms) charge radii of stable nuclei can be measured with electron scattering or atomic spectroscopy, and that for unstable nuclei can be deduced from the field shift of atomic spectra \cite{DeVries:1987atn,fricke2004nuclear,Angeli:2013epw}. The full charge distribution can only be extracted using electron scattering~\cite{DeVries:1987atn}. Either way, the Fermi function is based on direct experimental input, so the corresponding uncertainty is quantifiable. 

The shape factor $C(E_e)$ is a much smaller effect, but it depends on the CW current distribution $\rho_\text{cw}(\vec{x})$ which is not accessible experimentally,  other than with the decay itself. Generally, one may write
$\rho_\text{cw}(\vec{x})=\rho_\text{ch}(\vec{x})+\delta \rho(\vec{x})$.
%
%
Historically, the nuclear shell model calculations with one-body operators were used to compute $\delta \rho(\vec{x})$~\cite{Wilkinson:1993hx,Hardy:2004id}. The result was typically small, but no theory error was assigned to such a calculation. It is desirable to compute $\rho_\text{cw}$ with modern ab-initio methods and, maybe more importantly, corroborate a robust theory uncertainty. 


\subsubsection{Shape factor from isospin symmetry and nuclear charge radii\label{sec:rhocwsolution}}

A promising strategy for a model-independent  estimate of $C(E_e)$ was outlined in 
Ref.\cite{Seng:2022inj} and applied in Ref.\cite{Seng:2023cgl}. Isospin symmetry relates the CW distribution to linear combinations of the nuclear charge distributions across a nuclear isomultiplet. For superallowed transitions among $T=1$ states, the isospin structure of $J_W^\mu$ and $J_\text{em}^\mu$ in Eq.\eqref{eq:currents} implies the following relations:
\begin{align}
    &\rho_\text{cw}(r)=\rho_\text{ch,1}(r)+Z_0\left(\rho_\text{ch,0}(r)-\rho_\text{ch,1}(r)\right)
   =\rho_\text{ch,1}(r)+\frac{Z_{-1}}{2}\left(\rho_\text{ch,-1}(r)-\rho_\text{ch,1}(r)\right),\label{eq:rhocWCVC}\\
   &2Z_0\rho_\text{ch,0}(r)={Z_{-1}}\rho_\text{ch,-1}(r)+{Z_{1}}\rho_\text{ch,1}(r),
   \label{eq:rhochCVC}
\end{align}
where $Z_{T_z}$ is the atomic number of the nucleus with isospin quantum numbers $(1,T_z)$. 
%
%
\begin{figure}
    \centering
    \includegraphics[width=0.5\columnwidth]{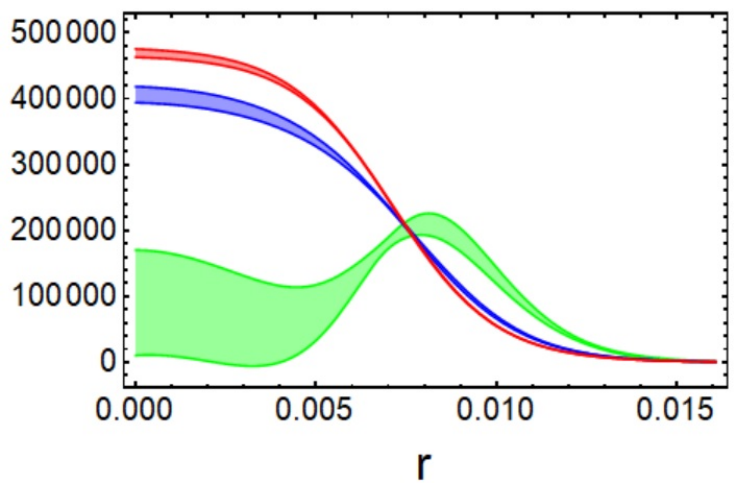}
    \caption{Nuclear charge distributions $\rho_\text{ch}(r)$ in atomic units for $^{22}$Mg (blue), $^{22}$Ne (red), and the corresponding CW distribution $\rho_\text{cw}(r)$ (green).}
    \label{fig:distributions}
\end{figure}
An application of this approach to the $^{22}$Mg $\to{}^{22}$Ne transition (we refer the reader to Ref.\cite{Seng:2023cgl} for a description of the parameters and uncertainties of charge distributions) is displayed in Fig. \ref{fig:distributions}. We observe that isospin symmetry predicts a stark difference between $\rho_\text{cw}(r)$ and any of the two charge distributions, i.e. $\delta\rho(r)$ is \textit{NOT} small.  A simple physical interpretation of this behavior is that only the protons in the outer shell can undergo the weak transition: a decay of those in the closed inert core is Pauli-suppressed as all neutron states are occupied. On the contrary, a virtual photon probes the entire nuclear charge. 

\begin{table*}[t]
\caption{\label{tab:RCW}Determinations of $\langle r_\text{cw}^2\rangle$ based on available data of nuclear charge radii for isotriplets in measured superallowed decays.  Notation: 123.12(234) means $123.12\pm 2.34$. Superscripts denote the source of data: Ref.\cite{Angeli:2013epw}$^a$, Ref.\cite{Li:2021fmk}$^b$, Ref.\cite{miller2019proton}$^c$, Ref.\cite{Bissell:2014vva}$^d$, Ref.\cite{Pineda:2021shy}$^e$, and Ref.\cite{Plattner:2023fmu}$^f$}
	\begin{centering}
	\begin{tabular}{c|c|c|c|c}
		\hline 
		$A$ & $\langle r_{\text{ch},-1}^2\rangle^{1/2}$ (fm) & $\langle r_{\text{ch},0}^2\rangle^{1/2}$ (fm) & $\langle r_{\text{ch},1}^2\rangle^{1/2}$ (fm) & $\langle r_{\text{cw}}^{2}\rangle^{1/2} $ (fm)\tabularnewline
		\hline 
		\hline 
		10 & $_{6}^{10}$C & $_{5}^{10}$B(ex) & $_{4}^{10}$Be: 2.3550(170)$^{a}$ & N/A\tabularnewline
		\hline 
		14 & $_{8}^{14}$O & $_{7}^{14}$N(ex) & $_{6}^{14}$C: 2.5025(87)$^{a}$ & N/A\tabularnewline
		\hline 
		18 & $_{10}^{18}$Ne: 2.9714(76)$^{a}$ & $_{9}^{18}$F(ex) & $_{8}^{18}$O: 2.7726(56)$^{a}$ & 3.661(72)\tabularnewline
		\hline 
		22 & $_{12}^{22}$Mg: 3.0691(89)$^{b}$ & $_{11}^{22}$Na(ex) & $_{10}^{22}$Ne: 2.9525(40)$^{a}$ & 3.596(99)\tabularnewline
		\hline 
		26 & $_{14}^{26}$Si & $_{13}^{26m}$Al: 3.130(15)$^f$ & $_{12}^{26}$Mg: 3.0337(18)$^{a}$ & 4.11(15)\tabularnewline
		\hline 
		30 & $_{16}^{30}$S & $_{15}^{30}$P(ex) & $_{14}^{30}$Si: 3.1336(40)$^{a}$ & N/A\tabularnewline
		\hline 
		34 & $_{18}^{34}$Ar: 3.3654(40)$^{a}$ & $_{17}^{34}$Cl & $_{16}^{34}$S: 3.2847(21)$^{a}$ & 3.954(68)\tabularnewline
		\hline 
		38 & $_{20}^{38}$Ca: 3.467(1)$^{c}$ & $_{19}^{38m}$K: 3.437(4)$^{d}$ & $_{18}^{38}$Ar: 3.4028(19)$^{a}$ & 3.999(35)\tabularnewline
		\hline 
		42 & $_{22}^{42}$Ti & $_{21}^{42}$Sc: 3.5702(238)$^{a}$ & $_{20}^{42}$Ca: 3.5081(21)$^{a}$ & 4.64(39)\tabularnewline
		\hline 
		46 & $_{24}^{46}$Cr & $_{23}^{46}$V & $_{22}^{46}$Ti: 3.6070(22)$^{a}$ & N/A\tabularnewline
		\hline 
		50 & $_{26}^{50}$Fe & $_{25}^{50}$Mn: 3.7120(196)$^{a}$ & $_{24}^{50}$Cr: 3.6588(65)$^{a}$ & 4.82(39)\tabularnewline
		\hline 
		54 & $_{28}^{54}$Ni: 3.738(4)$^{e}$ & $_{27}^{54}$Co & $_{26}^{54}$Fe: 3.6933(19)$^{a}$ & 4.28(11)\tabularnewline
		\hline 
		62 & $_{32}^{62}$Ge & $_{31}^{62}$Ga & $_{30}^{62}$Zn: 3.9031(69)$^{b}$ & N/A\tabularnewline
		\hline 
		66 & $_{34}^{66}$Se & $_{33}^{66}$As & $_{32}^{66}$Ge & N/A \tabularnewline
		\hline 
		70 & $_{36}^{70}$Kr & $_{35}^{70}$Br & $_{34}^{70}$Se & N/A\tabularnewline
		\hline 
		74 & $_{38}^{74}$Sr & $_{37}^{74}$Rb: 4.1935(172)$^{b}$ & $_{36}^{74}$Kr: 4.1870(41)$^{a}$ & 4.42(62)\tabularnewline
		\hline 
	\end{tabular}
		\par\end{centering}
	\end{table*}

The mean squared radii for $\rho_\text{ch}$ and $\rho_\text{cw}$ are defined according to
\begin{equation}
\langle r^2\rangle \equiv 4\pi \int_0^\infty dr r^2 \: r^2 \rho(r)~.
\end{equation}
Then, $\langle r_\text{cw}^2\rangle$ can be predicted from Eq.\eqref{eq:rhocWCVC}, provided two out of three nuclear charge radii are experimentally measured. With the currently available data, this can be done for $A=$18, 22, 26, 34, 38, 42, 50, 54 and 74 systems, see the last column in Tab.\ref{tab:RCW}. Due to the $Z$-enhanced isospin-breaking term in Eq.\eqref{eq:rhocWCVC}, the uncertainty in $\langle r_\text{cw}^2\rangle$ is in general much larger than that of the individual $\langle r_\text{ch}^2\rangle$. More importantly, the central value is significantly larger in most cases, in contradiction to older estimates~\cite{Wilkinson:1993hx,Hardy:2004id}. 


A simultaneous, fully data-driven evaluation of both $F(E_e)$ and $C(E_e)$ requires the information of at least two nuclear charge distributions within the nuclear isotriplet. 
This analysis was recently performed in Ref.~\cite{Seng:2023cgl}, and we report the results in Tab.~\ref{tab:final}. 
\begin{table}
	\begin{centering}		\begin{tabular}{c|c|c|c}
			\hline 
			Transition & $f_{\text{\text{new}}}$ & $f_{\text{HT}}$ & $\frac{f_{\text{new}}-f_{\text{HT}}}{f_{\text{new}}}$ (\%)\tabularnewline
			\hline 
			\hline 
			$^{18}$Ne$\rightarrow$$^{18}$F & $134.62(0)_{\text{dist}}(2)_\text{scr}(17)_{Q_{\text{EC}}}$ & $134.64(17)_{Q_{\text{EC}}}$ & $-0.01(0)_{\text{dist}}(2)_\text{scr}$\tabularnewline
			\hline 
			$^{22}$Mg$\rightarrow$$^{22}$Na & $418.27(2)_{\text{dist}}(7)_\text{scr}(13)_{Q_{\text{EC}}}$ & $418.35(13)_{Q_{\text{EC}}}$ & $-0.02(0)_{\text{dist}}(2)_\text{scr}$\tabularnewline
			\hline 
			$^{34}$Ar$\rightarrow$$^{34}$Cl & $3409.89(24)_{\text{dist}}(60)_\text{scr}(25)_{Q_{\text{EC}}}$ & $3410.85(25)_{Q_{\text{EC}}}$ & $-0.03(1)_{\text{dist}}(2)_\text{scr}$\tabularnewline
			\hline 
			$^{38}$Ca$\rightarrow$$^{38m}$K & $5327.49(39)_{\text{dist}}(98)_\text{scr}(31)_{Q_{\text{EC}}}$ & $5328.88(31)_{Q_{\text{EC}}}$ & $-0.03(1)_{\text{dist}}(2)_\text{scr}$\tabularnewline
			\hline 
			$^{42}$Ti$\rightarrow$$^{42}$Sc & $7124.3(58)_{\text{dist}}(14)_\text{scr}(14)_{Q_{\text{EC}}}$ & $7130.1(14)_{Q_{\text{EC}}}$ & $-0.08(8)_{\text{dist}}(2)_\text{scr}$\tabularnewline
			\hline 
			$^{50}$Fe$\rightarrow$$^{50}$Mn & $15053(18)_{\text{dist}}(3)_\text{scr}(60)_{Q_{\text{EC}}}$ & $15060(60)_{Q_{\text{EC}}}$ & $-0.04(12)_{\text{dist}}(2)_\text{scr}$\tabularnewline
			\hline 
			$^{54}$Ni$\rightarrow$$^{54}$Co & $21137(4)_{\text{dist}}(5)_\text{scr}(52)_{Q_{\text{EC}}}$ & $21137(57)_{Q_{\text{EC}}}$ & $+0.00(2)_{\text{rad}}(2)_\text{scr}$\tabularnewline
			\hline 
			$^{34}$Cl$\rightarrow$$^{34}$S & $1995.08(13)_{\text{dist}}(36)_\text{scr}(9)_{Q_{\text{EC}}}$ & $1996.003(96)_{Q_{\text{EC}}}$ & $-0.05(1)_{\text{dist}}(2)_\text{scr}$\tabularnewline
			\hline 
			$^{38m}$K$\rightarrow$$^{38}$Ar & $3296.32(22)_{\text{dist}}(63)_\text{scr}(15)_{Q_{\text{EC}}}$ & $3297.39(15)_{Q_{\text{EC}}}$ & $-0.03(1)_{\text{dist}}(2)_\text{scr}$\tabularnewline
			\hline 
			$^{42}$Sc$\rightarrow$$^{42}$Ca & $4468.53(340)_{\text{dist}}(91)_\text{scr}(46)_{Q_{\text{EC}}}$ & $4472.46(46)_{Q_{\text{EC}}}$ & $-0.09(8)_{\text{dist}}(2)_\text{scr}$\tabularnewline
			\hline 
			$^{50}$Mn$\rightarrow$$^{50}$Cr & $10737.9(117)_{\text{dist}}(23)_\text{scr}(5)_{Q_{\text{EC}}}$ & $10745.99(49)_{Q_{\text{EC}}}$ & $-0.08(11)_{\text{dist}}(2)_\text{scr}$\tabularnewline
			\hline 
			$^{54}$Co$\rightarrow$$^{54}$Fe & $15769.4(24)_{\text{dist}}(34)_\text{scr}(27)_{Q_{\text{EC}}}$ & $15766.8(27)_{Q_{\text{EC}}}$ & $+0.02(2)_{\text{dist}}(2)_\text{scr}$\tabularnewline
			\hline 
			$^{74}$Rb$\rightarrow$$^{74}$Kr & $47326(128)_{\text{dist}}(12)_\text{scr}(94)_{Q_{\text{EC}}}$ & $47281(93)_{Q_{\text{EC}}}$ & $+0.10(27)_{\text{dist}}(3)_\text{scr}$\tabularnewline
			\hline 
		\end{tabular}
		\par	   
	\end{centering}

	\caption{\label{tab:final}Comparison between new and old results of $f$. The three sources of uncertainty are from charge distributions in the Fermi function and the shape factor (dist), screening correction (scr) and the decay $Q$-value ($Q_{EC}$), respectively. Numerical values from Ref.\cite{Seng:2023cgl}.}
	
\end{table}
%
We observe that adopting this new approach to determine the statistical rate function, the shift in the central values is not negligible, and neither is the associated uncertainty. 
It has to be noted that at the needed precision level the shape factor is governed by the value of the weak radius and is insensitive to finer details of the weak distribution. 
One sees that for
$A$=10, 14, 30, 46 and 62 systems, the addition of one single charge radius measurement will permit to predict $\langle r_\text{cw}\rangle$ and evaluate the shape factor with the data-driven uncertainty. 

We conclude this subsection by noting that the $Z$-enhancement in  Eq.\eqref{eq:rhocWCVC} responsible for a large deviation of $\rho_\text{cw}$ from $\rho_\text{ch}$ and respective radii, would also lead to an enhanced sensitivity to isospin symmetry breaking, neglected until now. To check the validity of this assumption, it is crucial to have data on all three charge radii in the isotriplet to be able to use Eq.\eqref{eq:rhochCVC} for a quantitative test.
To this end, we point out that a possible future experimental program aiming at high-precision extraction of charge radii for all nuclear isotopes listed in Tab.\ref{tab:RCW} would have a significant impact. 
This can be achieved with the X-ray spectroscopy of muonic atoms at PSI in the case of light stable isotopes \cite{ohayon2023precision}. Unstable  radioisotopes can be produced at rare isotope facilities such as ISOLDE~\cite{ISOLDE}, FRIB~\cite{FRIB}, GSI/FAIR~\cite{GSI-FAIR}, RIKEN~\cite{RIKEN} and TRIUMF~\cite{TRIUMF}. For instance, a possibility to produce  some long-lived radioisotopes at ISOLDE, transport them to PSI where their charge radii can be measured was put forward in Ref.\cite{isolde-psi}. 
We refer the reader to a recent review on nuclear charge radii and pertinent measurement techniques in Ref.\cite{Nortershauser:2023jwa}.


 \subsection{Fermi matrix element and ISB correction} 

We next turn to the second ingredient in Eq.\eqref{eq:treeME}, the Fermi matrix element:
$M_F\equiv\langle \phi_f|\tau_+|\phi_i\rangle$.
In the isospin limit, its value is exactly known,  $M_F^0=\sqrt{2}$ for transitions within an isotriplet. However, in presence of ISB, this value receives a correction,
\begin{equation}
M_F^2=(M_F^{0})^2(1-\delta_\text{C})~.
\end{equation}
It primarily originates from the Coulomb repulsion between the protons in the nucleus (hence the subscript C), but receives further, generally smaller contributions from other charge-dependent nuclear forces. The size of $\delta_\text{C}$ generally ranges from 0.1\% to 1\%, 
thus playing a crucial role in the test of the $\mathcal{F}t$-value universality (in other words, CVC). Theoretical computations of $\delta_\text{C}$ have been a classical problem in nuclear physics since 1950's~\cite{MacDonald:1958zz}.

\subsubsection{Historical approaches}

In principle, computing the full Fermi matrix element $M_F$ directly in the presence of ISB interactions and comparing it to the isospin limit $M_F^0$ straightforwardly gives $\delta_\text{C}$. In practice, while straightforward, this calculation is complicated, not least due to the fact that no single model gives a consistent description for all nuclei with $10\leq A\leq74$. The outcome of several selected approaches for seven transitions is summarized in Tab.\ref{tab:ISB}. One observes a significant spread between different model predictions, raising questions about the uncertainty of $\delta_\text{C}$. Given this discrepancy, Hardy and Towner advocated the CVC hypothesis test~\cite{Towner:2010bx} to discriminate between models. Assuming that no BSM contributions are present, all $\mathcal{F}t$-values must align. Each model calculation of 
$\delta_\text{C}$ is then combined with other corrections in Eq.\eqref{eq:Ft}, and a constant fit to the 15 most precise transitions is performed. The model that delivers the lowest $\chi^2$ is then confirmed, while all others are refuted. The only model that passed this test was the nuclear shell-model with Woods-Saxon (WS) potential which was adjusted in each isotriplet to reproduce the available ISB-sensitive observables~\cite{Towner:2007np}. It has long been considered the most appropriate and reliable tool, and included in the superallowed beta decay reviews~\cite{Hardy:2008gy,Hardy:2014qxa,Hardy:2020qwl}.

\begin{table*}[t]
\caption{\label{tab:ISB}Results for $\delta_\text{C}$ in the nuclear shell model with the Woods-Saxon (WS) potential~\cite{Hardy:2020qwl}, the density functional theory (DFT)~\cite{Satula:2016hbs}, the Hartree-Fock (HF) calculation~\cite{Ormand:1995df}, the random phase
approximation (RPA)~\cite{Liang:2009pf}, and the isovector monopole-resonance model (IVMR)~\cite{Auerbach:2008ut}.}
	\begin{centering}
		\begin{tabular}{c|c|c|c|c|c}
			\hline 
			Transitions & \multicolumn{1}{c}{} & \multicolumn{1}{c}{} & \multicolumn{1}{c}{$\delta_{\text{C}}$} & \multicolumn{1}{c}{(\%)} & \tabularnewline
			\cline{2-6} 
			& WS & DFT & HF & RPA & IVMR \tabularnewline
			\hline 
			\hline 
			$^{26m}\text{Al}\rightarrow^{26}\!\!\text{Mg}$ & 0.310 & 0.329 & 0.30 & 0.139 & 0.08\tabularnewline
			\hline 
			$^{34}\text{Cl}\rightarrow^{34}\!\!\text{S}$ & 0.613 & 0.75 & 0.57 & 0.234 & 0.13 \tabularnewline
			\hline 
			$^{38m}\text{K}\rightarrow^{38}\!\!\text{Ar}$ & 0.628 & 1.7 & 0.59 & 0.278 & 0.15\tabularnewline
			\hline 
			$^{42}\text{Sc}\rightarrow^{42}\!\!\text{Ca}$ & 0.690 & 0.77 & 0.42 & 0.333 & 0.18 \tabularnewline
			\hline 
			$^{46}\text{V}\rightarrow^{46}\!\!\text{Ti}$ & 0.620 & 0.563 & 0.38 & / & 0.21 \tabularnewline
			\hline 
			$^{50}\text{Mn}\rightarrow^{50}\!\!\text{Cr}$ & 0.660 & 0.476 & 0.35 & / & 0.24 \tabularnewline
			\hline 
			$^{54}\text{Co}\rightarrow^{54}\!\!\text{Fe}$ & 0.770 & 0.586 & 0.44 & 0.319 & 0.28\tabularnewline
			\hline 
		\end{tabular}
		\par\end{centering}
\end{table*}

Ref.\cite{Towner:2007np} expresses $M_F$ as a nuclear matrix element of a one-body operator,
\begin{equation}
M_F=\sum_{\alpha\beta}\langle \phi_f|a_{n,\alpha}^\dagger a_{p,\beta}|\phi_i\rangle \langle n,\alpha|\tau_+|p,\beta\rangle~.~\label{eq:MFshell}
\end{equation}
Above, $a_{n,\alpha}^\dagger$ is an operator that creates a neutron in the state $\alpha$ and $a_{p,\beta}$ annihilates a proton in the state $\beta$. Here we distinguish protons and neutrons explicitly to avoid confusion. The one-body matrix element is parametrized as
\begin{equation}
\langle n,\alpha|\tau_+|p,\beta\rangle\equiv \delta_{\alpha\beta}r_\alpha,
\label{eq:radialint}
\end{equation}
with $r_\alpha$ the radial integral that equals 1 in the isospin limit. Inserting a complete set of $A-1$-nucleon states $\{|\pi\rangle\}$ into Eq.\eqref{eq:MFshell}, and allowing $r_\alpha$ to depend on $\pi$,  Eq.\eqref{eq:MFshell} is recast as
\begin{equation}
M_F\rightarrow \sum_{\pi,\alpha}\langle \phi_f|a_{n,\alpha}^\dagger |\pi\rangle \langle \pi|a_{p,\alpha}|\phi_i\rangle r_\alpha^\pi~.
\end{equation}
Because the Coulomb interaction is long-range, the number of the shell-model configurations that need to be accounted for is large, and the ISB correction is split as $\delta_\text{C}=\delta_\text{C1}+\delta_\text{C2}$, with the two terms defined as follows: 
\begin{enumerate}
    \item 
$\delta_\text{C1}$ is the isospin-mixing correction and stems from breaking the equality $\langle \pi|a_{p,\alpha}|\phi_i\rangle^*=\langle \phi_f|a_{n,\alpha}^\dagger|\pi\rangle$ due to ISB.
    \item 
$\delta_\text{C2}$ is the radial mismatch correction: even in the absence of isospin mixing, the radial integral can differ from 1.
\end{enumerate}
In practice, $\delta_\text{C2}$ is numerically larger; it is computed using the shell model, supplemented by the experimental information of spectroscopic factors measured in neutron pick-up direct reactions that provide guidance on which orbitals $\alpha$ to be included in the model space. On the other hand, $\delta_\text{C1}$ depends on spectroscopic amplitudes involving non-analog $0^+$ states; they are again computed using the shell model, with the model parameters tuned to maximally reproduce the isobaric multiplet
mass equation (IMME)~\cite{Lam:2013bhc}, as well as the excitation energies of the non-analog $0^+$ states. 

Several drawbacks of this formalism were pointed out by Miller and Schwenk~\cite{Miller:2008my,Miller:2009cg}. For instance, the parameterization of the single-particle matrix element in Eq.\eqref{eq:radialint} that assumes $\alpha=\beta$, results in the violation of the standard isospin commutation relations. The inclusion of $\alpha\neq \beta$ (i.e., radial excitation) contributions could lead to a significant decrease of $\delta_\text{C}$~\cite{Miller:2009cg}. Another criticism is that the splitting $\delta_\text{C}=\delta_\text{C1}+\delta_\text{C2}$ is model dependent and unnatural in a proper isospin formalism, therefore the WS shell-model calculation may be prone to doublecounting.
The CVC-based selection is also questionable: rather than failure of a nuclear model, a non-constant $\mathcal{F}t$-value could be interpreted as the BSM signal. 


In summary, the discussion in the nuclear theory community about the correctness and accuracy of the nuclear shell model calculation with WS potential appears to still be open. Modern ab-initio calculations are expected in the near future. Nonetheless, the question of a model-independent assessment of the uncertainty of any single calculation remains. Considering that the consistency across the superallowed transition chart is the basis for the $V_{ud}$ extraction, a model-independent approach to the problem is needed.

\subsubsection{Towards data-driven $\delta_\text{C}$}

Refs.\cite{Seng:2022epj,Seng:2023cvt} approached this task by combining the perturbation theory framework of  Refs.\cite{Miller:2008my,Miller:2009cg} with the ideas of Refs.\cite{Damgaard:1969yyx,Auerbach:2008ut}.
The full Hamiltonian can be split as
$H=H_0+V$,
with $H_0$ isospin-symmetric and $V$ the ISB parts, respectively. Behrens-Sirlin-Ademollo-Gatto theorem~\cite{Behrends:1960nf,Ademollo:1964sr} states that $\delta_\text{C}$ scales in ISB as $\mathcal{O}(V^2)$. Assuming $V$ to be  predominantly isovector, as suggested by the data on IMME, the Wigner-Eckart theorem gives
\begin{equation}
\delta_\text{C}=\frac{1}{3}\sum_a\frac{|\langle a;0||V||g;1\rangle|^2}{(E_{a,0}-E_{g,1})^2}+\frac{1}{2}\sum_{a\neq g}\frac{|\langle a;1||V||g;1\rangle|^2}{(E_{a,1}-E_{g,1})^2}-\frac{5}{6}\sum_a\frac{|\langle a;2||V||g;1\rangle|^2}{(E_{a,2}-E_{g,1})^2}+\mathcal{O}(V^3)~,\label{eq:deltaCPT}
\end{equation}
where $\langle a;0||V||g;1\rangle $ are reduced matrix elements of $V$ between the ground state (g) $J^P=0^+$ isotriplet $|g;1,T_z\rangle$ and a general intermediate state $|a;T',T_z'\rangle$, with $a$ representing collectively all the non-isospin quantum numbers. The fact that only excited states with $\{a,T'\}\neq \{g,1\}$ contributes to the expression above (in contrast to IMME which depend on ground-state matrix elements) suggests a similarity to the isospin-mixing effect represented by $\delta_\text{C1}$ in the Hardy-Towner formalism. However, we want to stress that Eq.\eqref{eq:deltaCPT} is a rigorous result from quantum mechanical perturbation theory, so it should contain all the contributions at the order $\mathcal{O}(V^2)$, both from $\delta_\text{C1}$ and $\delta_{C2}$, as well as contributions that are intrinsically missing in the Hardy-Towner formalism. 
%

Following Refs.\cite{Damgaard:1969yyx,Auerbach:2008ut}, Ref.\cite{Seng:2022epj} associated the ISB potential with the isovector part of the Coulomb potential. The one-body isovector Coulomb potential, taken for the uniform sphere inside which all the protons reside, is identified with the isovector monopole operator 
\begin{equation}
    \vec{M}^{(1)}=\sum_i r_i^2\vec{T}_i~,~M_0^{(1)}=M_z^{(1)}~,~M_{\pm 1}^{(1)}=\mp (M_x^{(1)}\pm i M_y^{(1)})/\sqrt{2}~.
\end{equation}
Sandwiching this operator between nuclear states gives the isovector nuclear radii
which are measurable quantities.
As a result, ISB-sensitive combinations of electroweak nuclear radii could be constructed,
\begin{eqnarray}
    \Delta M_A^{(1)} &\equiv & 
    -\langle r_\text{cw}^2\rangle + \left(\frac{N_1}{2}\langle r_{n,1}^2\rangle-\frac{Z_1}{2}\langle r_{p,1}^2\rangle\right)
\nonumber\\
    \Delta M_B^{(1)}&\equiv &\frac{1}{2}(Z_1\langle r_{\text{ch,1}\rangle}^2+Z_{-1}\langle r_{\text{ch,-1}}^2\rangle)-Z_0\langle r_{\text{ch,0}}^2\rangle~.
    \label{eq:dMAdMB}
\end{eqnarray}
Above, $\langle r_\text{cw}^2\rangle $ and $\langle r_\text{ch}^2\rangle$ are the nuclear mean squared CW and charge radii already introduced in Sec.\ref{sec:rhocwsolution}. 
The new ingredient, $\langle r_n^2\rangle-\langle r_p^2\rangle$, is the difference between the mean squared neutron and proton distribution radius in a nucleus, which is related to the neutron skin. One may easily show that both $\Delta M_{A,B}^{(1)}$ vanish in the isospin limit (in fact, $\Delta M_B^{(1)}$ is precisely the amount by which the isospin symmetric relation in the second line of Eq.\eqref{eq:rhocWCVC} is violated). 

The expressions for $\Delta M_{A,B}^{(1)}$ similar to Eq.\eqref{eq:deltaCPT} are reported in Ref.\cite{Seng:2022epj}, and we do not display them here. In the simplest case of a single term in the sum over nuclear intermediate states in Eq.\eqref{eq:deltaCPT}, low-lying isovector monopole resonance dominance picture $\Delta M_{A,B}^{(1)}$ and $\delta_\text{C}$ are straightforwardly related. This allowed for first numerical estimates, and an observation of the hierarchy $\Delta M_{A}^{(1)}=O(V)$, $\delta_\text{C}=O(V^2)$ and $\Delta M_{B}^{(1)}=O(V^3)$, although formally $\Delta M_{B}^{(1)}$ is of the same order in ISB as $\Delta M_{A}^{(1)}$. This hieararchy is worth investigating further, keeping in mind the isospin relations used in Sec.\ref{sec:rhocwsolution}.

In the general case, due to the infinite sum in Eq.\eqref{eq:deltaCPT}, the relation between the ISB combinations of nuclear radii and the ISB correction to $M_F$ is not straightforward. 
Ref.\cite{Seng:2023cvt} circumvented this complication by defining a set of generating functions,
\begin{equation}
    F_{T_z}(\omega)=-\langle g;1,T_z|(M_{-1}^{(1)})^\dagger G(\omega)M_{-1}^{(1)}|g;1,T_z\rangle+\frac{|\langle g;1,T_z-1|M_{-1}^{(1)}|g;1,T_z\rangle|^2}{E_{g,1}-\omega}~,
\end{equation}
where $G(\omega)\equiv 1/(H_0-\omega)$ is the nuclear Green's function of the isospin-symmetric system, with $\omega$ a free energy parameter. If one assumes the dominance of the Coulomb interaction in $V$, then $\Delta M_{A,B}^{(1)}$ and $\delta_\text{C}$ are related to the generating function and its derivative at $\omega=E_{g,1}$, respectively. 
Therefore, if $F_{T_z}(\omega)$ as a function of $\omega$ can be computed in a given nuclear theory framework, both $\Delta M_{A,B}^{(1)}$ and $\delta_\text{C}$ can be predicted with a fully-correlated theory uncertainties, and the former can be directly constrained by experiment. This idea thus offers a promising framework for a data-driven analysis of $\delta_\text{C}$.  

\subsubsection{Experimental opportunities}

We briefly discuss the experimental inputs needed to determine the observables in Eq.\eqref{eq:dMAdMB}, starting from $\Delta M_B^{(1)}$ that requires only the nuclear charge radii, which are the most commonly measured electroweak radii. If isospin symmetry was assumed, then two known charge radii in a nuclear isotripet are sufficient to fix the whole system, which is the underlying principle in the discussion of Sec.\ref{sec:rhocwsolution}; however, now we are probing the ISB effect so a third radii measurement is needed. From Tab.\ref{tab:RCW}, we see that only the $A=38$ system has all charge radii measured, and the resulting ISB observable is consistent with zero: $\Delta M_B^{(1)}=0.00(12)_\text{Ca}(52)_\text{K}(14)_\text{Ar}$. The current limitation is the $^{38m}$K charge radius, and an improvement of its precision by a factor $2$ may start to discriminate among nuclear models~\cite{Seng:2022epj}. In the meantime, an addition of one single radius measurement in $A=$18, 22, 34, 42, 50, 54 and 74 systems will activate the ISB analysis through $\Delta M_B^{(1)}$. 

$\Delta M_A^{(1)}$ combines weak nuclear radii that are more difficult to measure. On the other hand, much lesser experimental precision than for $\Delta M_B^{(1)}$ is required, and a percent-level determination of these radii will start to discriminate models~\cite{Seng:2022epj}. The neutron skin of neutron-rich nuclei is an important experimental constraint on the nuclear equation of state~\cite{Brown:2000pd,Horowitz:2014bja,Kumar:2020ejz}, and can be measured with parity-violating electron scattering (PVES). The recent PREX and CREX experiments achieved a percent-level determination of the neutron radii of $^{208}$Pb and $^{48}$Ca~\cite{PREX:2021umo,CREX:2022kgg}, and will further be improved with MREX experiment~\cite{Becker:2018ggl}. While for neutron-rich nuclei the neutron skin is primarily generated by the symmetry energy, the neutron skin of symmetric nuclei is a pure ISB effect. A determination of the $^{12}$C neutron skin at Mainz is feasible~\cite{Becker:2018ggl,Koshchii:2020qkr}, opening the possibility of measuring neutron skins of stable superallowed daughters. A direct determination of $\langle r_\text{cw}^2\rangle$ is challenging, as it requires precise measurements of small recoil effects in nuclear beta decays. This may become possible with the next generation experiments~\cite{Leach:2021bvh,Carney:2022pku}.

\section{Introducing one-loop radiative corrections}

\begin{figure}
    \centering
    \includegraphics[width=0.8\columnwidth]{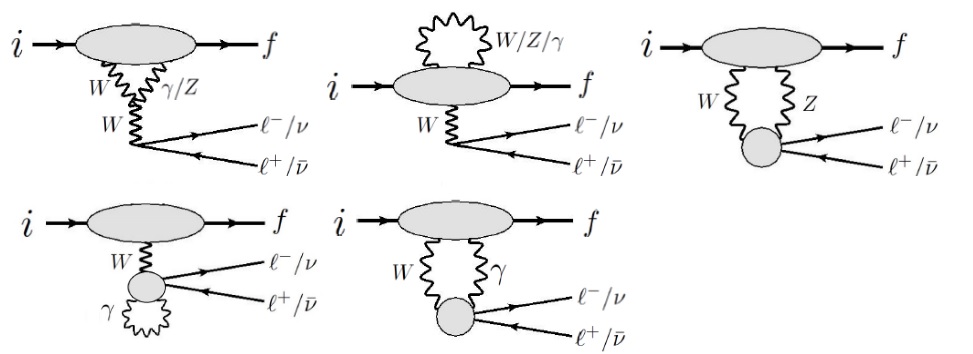}
    \caption{One-loop corrections in generic semileptonic beta decays.   }
    \label{fig:loops}
\end{figure}
The preceding Section was dedicated to tree-level corrections, and we move on to one-loop RC. The division is somewhat ambiguous: $\delta_\text{C}$ largely arises due to Coulomb interaction among the protons in the nucleus, and in the language of Feynman diagrams may correspond to an electromagnetic correction to the weak vertex. Similarly, the infrared (IR)-sensitive part of the $\gamma W$-box diagram is included in the Fermi function. This separation strategy is largely based on
the current algebra formalism by Sirlin~\cite{Sirlin:1977sv,Seng:2021syx}.

We start by introducing some terminology. Electromagnetic corrections to the differential decay rate are of two kinds: (1) Loop corrections where virtual photons are exchanged, and (2) Bremsstrahlung corrections where an extra real photon is emitted. These two corrections have to be considered simultaneously to ensure that the outcome is IR-finite~\cite{Bloch:1937pw,Yennie:1961ad,Kinoshita:1958ru,Kinoshita:1962ur,Lee:1964is}. These corrections modify the differential decay rate as
\begin{equation}
\frac{d\Gamma}{dE_e}\rightarrow F(E_e)(1+\delta_\text{out}(E_e))(1+\delta_\text{in}(E_e))\frac{d\Gamma}{dE_e}~.
\end{equation}
The first and largest effect is $F(E_e)$, the Fermi function introduced in the previous section. The second term $\delta_\text{out}(E_e)$, the outer correction, entails all IR-sensitive RC not included in $F(E_e)$.
It can be computed order-by-order in QED, and the first result is the $\mathcal{O}(\alpha)$ correction obtained by Sirlin~\cite{Sirlin:1967zza}: 
\begin{eqnarray}
\delta_\text{out}(E_e)&=&\frac{\alpha}{2\pi}\left\{3\ln\frac{m_p}{m_e}-\frac{3}{4}+4\left(\frac{1}{\beta}\tanh^{-1}\beta-1\right)\left(\ln\frac{2(E_0-E_e)}{m_e}+\frac{E_0-E_e}{3E_e}-\frac{3}{2}\right)\right.\nonumber\\
&&\left.-\frac{4}{\beta}\text{Li}_2\left(\frac{2\beta}{1+\beta}\right)+\frac{1}{\beta}\tanh^{-1}\beta\left(2+2\beta^2+\frac{(E_0-E_e)^2}{6E_e^2}-4\tanh^{-1}\beta\right)\right\}~.
\end{eqnarray}
Corrections to order $Z\alpha^2$, $Z^2\alpha^3$ and $\alpha^2$ were included later~\cite{Sirlin:1987sy,Sirlin:1986cc,Towner:2007np}. The Fermi function and the outer correction are functions of $E_e$ that affect the beta decay spectrum. 
All the remaining RC are collectively denoted as the inner correction $\delta_\text{in}(E_e)$ which depends on hadron/nuclear structure, and we explicitly indicate its energy dependence. 
While the Bremsstrahlung contributions are easily calculable, computing various one-loop diagrams involving virtual gauge bosons is more challenging. Many of them are identical to those which appear in muon decay $\mu\rightarrow \nu_\mu e\bar{\nu}_e$, and are absorbed into the renormalized Fermi constant $G_F$ obtained from the muon lifetime~\cite{MuLan:2012sih}. The remaining loop diagrams can be systematically analyzed with current algebra~\cite{Sirlin:1977sv,Seng:2021syx}. Restricting our consideration to superallowed decays, they lead to the following form of the inner correction:
\begin{equation}
\delta_\text{in}(E_e)=\Delta_R^U+2\Box_{\gamma W}^\text{nucl}~.\label{eq:Sirlin}
\end{equation}
The first term in the equation above, $\Delta_R^U=0.01709(10)$, represents a universal piece in all nuclear beta decays. It consists of the analytically-calculable weak RC, perturbative Quantum Chromodynamics (pQCD) corrections, resummed leading QED logarithms, and the most important $\mathcal{O}(\alpha^2)$ corrections~\cite{Czarnecki:2004cw}. Our focus is on the second term, where $\Box_{\gamma W}^\text{nucl}$
represents the $\gamma W$-box diagram, i.e. the last diagram in Fig.\ref{fig:loops}, evaluated on nuclei. 


\subsection{\label{sec:Boxn}Factorization and the single-nucleon box diagram}


Historically, 
a factorized form that separates the nucleus-independent and dependent pieces has been adopted,
\begin{equation}
1+\delta_\text{in}= 1+\Delta_R^U+2\Box_{\gamma W}^n+2(\Box_{\gamma W}^\text{nucl}-\Box_{\gamma W}^n)\approx (1+\Delta_R^V)(1+\delta_\text{NS})~,
\end{equation}
where $\Box_{\gamma W}^n$ is the $\gamma W$-box diagram correction to the free neutron decay. Here, we have defined $\Delta_R^V\equiv 1+\Delta_R^U+2\Box_{\gamma W}^n$ and $\delta_\text{NS}\equiv 2(\Box_{\gamma W}^\text{nucl}-\Box_{\gamma W}^n)$ which appear in the master formulas Eqs.\eqref{eq:master} and \eqref{eq:Ft}, respectively. 

Until 2018, the nucleus-independent RC $\Delta_R^V$ was believed to be the primary source of uncertainty in $V_{ud}$. The recent theory progress is summarized in a recent review \cite{Gorchtein:2023srs}. Computing $\Delta_R^V$ entails to calculate the loop integral in the free-nucleon box diagram $\Box_{\gamma W}^n$, with contributions from loop momenta at all scales, with a controlled precision. At large loop momenta pQCD prevails, the integral is process-independent and is perturbatively calculable~\cite{Baikov:2010iw,Baikov:2010je}. Below the perturbative regime, however, non-perturbative inputs are necessary. Among them, the contribution of the nucleon ground state propagating between the electromagnetic and weak vertices (Born contribution), is fixed by the well-determined nucleon axial and magnetic Sachs form factors~\cite{Ye:2017gyb,Lorenz:2012tm,Lorenz:2014yda,Lin:2008uz,Lin:2021umk,Lin:2021umz,Bhattacharya:2011ah}. The main theory uncertainty thus stems from the interpolation between the two extremes and is associated with multi-hadron intermediate states. Early works on $\Box_{\gamma W}^n$ can be traced back to Marciano and Sirlin~\cite{Marciano:1985pd,Marciano:2005ec}, where different interpolation strategies were used to mimic the multi-hadron contributions. 
The dispersion relation analysis of $\Box_{\gamma W}^n$ in 2018~\cite{Seng:2018yzq}, a method applied earlier to the $\gamma Z$-box diagram in PVES~\cite{Gorchtein:2008px}, reformulated the problem in a way that allows one to constrain the multi-hadron contributions with the input from neutrino scattering data or lattice QCD calculations. This dispersive evaluation was confirmed by a series of subsequent works~$\Box_{\gamma W}^n$~\cite{Seng:2018qru,Seng:2020wjq,Shiells:2020fqp,Czarnecki:2019mwq,Hayen:2020cxh}. In a recent review on neutron beta decay~\cite{Gorchtein:2023srs}, these results were combined into a recommended value of $\Delta_R^V$,
\begin{equation}
\overline{(\Delta_R^V)}_\text{DR}=0.02479(21)~.\label{eq:DeltaRVDR}
\end{equation}
This year, the first direct lattice QCD calculation appeared~\cite{Ma:2023kfr},
\begin{equation}
    (\Delta_R^V)_\text{lat}=0.02439(19)~,\label{eq:DeltaRVlat}
\end{equation}
showing a slight 1.4$\sigma$ tension with the DR result, which is worth investigating further. 

\section{\label{sec:nucleargammaW}Nuclear $\gamma W$-box in the dispersive formalism}

In this Section, we outline the rigorous theory framework for the nuclear $\gamma W$-box and $\delta_\text{NS}$. It largely follows Ref.\cite{Seng:2022cnq}, but a more detailed analysis of the $E_e$-dependence, not explicitly given in that reference, is presented here for the first time. 

The non-trivial part of the $\gamma W$-box diagram that gives rise to the inner correction reads
\begin{equation}
\Box_{\gamma W}(E_e)=\frac{e^2}{M_F}\mathfrak{Re}\int\frac{d^4q}{(2\pi)^4}\frac{M_W^2}{M_W^2-q^2}\frac{\left[Q^2+M\nu\frac{p\cdot q m_e^2-p_e\cdot q p\cdot p_e}{M^2m_e^2-(p\cdot p_e)^2}\right]T_3(\nu,Q^2)}{[(p_e-q)^2-m_e^2+i\varepsilon](q^2+i\varepsilon)M\nu}~,
\label{eq:gWboxGen}
\end{equation}
with $\nu=p\cdot q/M=q_0$ the loop photon energy in the nuclear rest frame, and $Q^2=-q^2$ its virtuality. We suppress the superscript ``nucl'' or ``$n$'' so it may refer to either case. The central quantity here is the invariant amplitude $T_3(\nu,Q^2)$ related to a time-ordered product of the hadronic EM and the CW current:
\begin{align}
T_3=-\frac{2M\nu}{\q}\sum_X\Bigg[\frac{\langle \phi_f|J_\text{em}^x(\vec{q})|X\rangle\langle X|J_{W5}^{\dagger y}(-\vec{q})|\phi_i\rangle}{\nu_X-\nu-i\varepsilon}
+\frac{\langle \phi_f|J_{W5}^{\dagger y}(-\vec{q})|X\rangle\langle X|J_\text{em}^x(\vec{q})|\phi_i\rangle}{\nu_X+\nu-i\varepsilon}\Bigg],
\label{eq:T3GF}
\end{align}
where $\nu_X\equiv E_X-M$,  $\vec{q}=\q \hat{z}$, and $J_{W5}$ stands for the axial part of the charged weak current. The effect of the time-ordered product is encoded in the Green's function, $G(\omega)= 1/(H_0-\omega)$ which counts all possible hadronic intermediate states. 
We note that in the above expressions we explicitly neglect nuclear recoil and work in the exact forward kinematics. Since the $\gamma W$-box is a radiative correction $\sim\alpha/\pi\sim10^{-3}$, including the recoil $\sim10^{-3}$ on top of it would exceed our precision goal of $10^{-4}$. Accordingly, it suffices to take $M_i\approx M_f\approx M$ in the calculation of the box diagram, and only account for the finite mass difference $M_i-M_f$ when convoluting the energy-dependent box diagram with the decay spectrum.

\begin{figure}
    \centering
    \includegraphics[width=0.5\columnwidth]{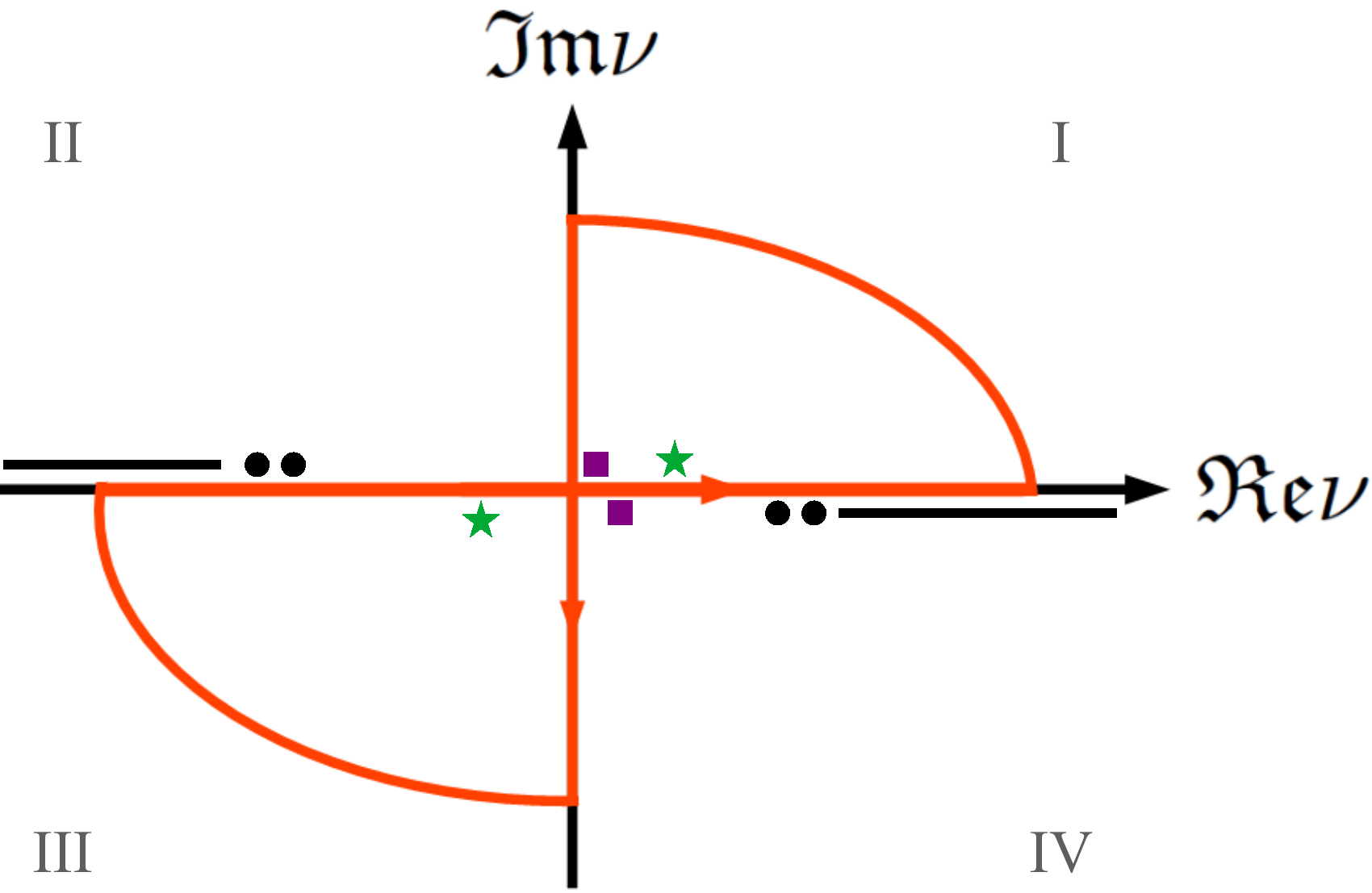}
    \caption{Contour in the complex $\nu$-plane. Black dots and lines: Poles and cuts from the photon and W propagator, and states with $M_X>M$. Purple boxes: Poles from the electron propagator. Green stars: Poles from states with $M_X<M$.  }
    \label{fig:contour}
\end{figure}

To evaluate $\Box_{\gamma W}$, we integrate over $\nu=q_0$ using Wick rotation following the contour in Fig.\ref{fig:contour}. For that, we locate and classify the singularities of the integrand:
\begin{enumerate}
    \item The photon and $W$-propagators give rise to poles in the  quadrants II and IV.
    \item The electron propagator has two poles: $\nu=E_e+\sqrt{|\vec{p}_e-\vec{q}|^2+m_e^2}-i\varepsilon$ is in the quadrant IV, while $\nu=E_e-\sqrt{|\vec{p}_e-\vec{q}|^2+m_e^2}+i\varepsilon$ can be in the  quadrant I or II.
    \item Intermediate states in $T_3$ with $M_X>M$: poles and cuts in the quadrant II and IV.
    \item If $\phi_i$ or $\phi_f$ is an excited state, 
    intermediate states in $T_3$ with $M_X<M$ are possible, which would lead to poles in the quadrant I or III. This possibility was recently pointed out by M. Drissi, M. Gennari and P. Navratil~\cite{Gennari}. Details of this contribution will be given in Subsection~5.4 for the first time.
\end{enumerate}
With these, performing the integral with the help of Wick rotation leads to three terms:
\begin{equation}
    \Box_{\gamma W}=(\Box_{\gamma W})_\text{Wick}+(\Box_{\gamma W})_{\text{res},e}+(\Box_{\gamma W})_{\text{res},T_3}\,.\label{eq:3terms}
\end{equation}
The first term is the Wick-rotated term where $\nu$ is substituted by $i\nu_E$. The second term picks up the pole of the electron propagator in the first quadrant when $E_e>\sqrt{|\vec{p}_e-\vec{q}|^2+m_e^2}$. The third term picks up the pole of $T_3$ in the first and third quadrant contributed by $M_X<M$ intermediate states. 
The third term in Eq.\eqref{eq:3terms} is a pure nuclear effect and exhibits a singular behavior in $E_e$ as will be discussed later. Meanwhile, as the first two terms are regular functions of $E_e$, we may expand their sum as:
\begin{equation}
    (\Box_{\gamma W})_\text{Wick}+(\Box_{\gamma W})_{\text{res},e}=\boxminus_0+\boxminus_1E_e+\mathcal{O}(E_e^2)~.\label{eq:regularexpand}
\end{equation}
A simple dimensional analysis suggests that the linear term $\boxminus_1 E_e$ is possibly relevant only for nuclear decays, and  
we finally can re-express $\delta_\text{NS}$ as
\begin{equation}
\delta_\text{NS}\approx 2(\boxminus_0^\text{nucl}-\boxminus_0^n)+2\boxminus_1^\text{nucl}\langle E_e\rangle+2\langle (\Box_{\gamma W}^\text{nucl}(E_e))_{\text{res},T_3}\rangle,\label{eq:deltaNSexpand}
\end{equation}
where the energy-average of a generic function $f(E_e)$ is defined as
\begin{equation}
\langle f(E_e)\rangle\equiv \frac{\int_{m_e}^{E_0}dE_e |\vec{p}_e|E_e(E_0-E_e)^2F(E_e)f(E_e)}{\int_{m_e}^{E_0}dE_e|\vec{p}_e|E_e(E_0-E_e)^2F(E_e)}~.
\end{equation}

\subsection{Dispersive representation of the regular terms}

We start with the $E_e$-regular terms. The first two expansion coefficients in Eq.\eqref{eq:regularexpand}  read~\cite{Seng:2022cnq}:
\begin{eqnarray}
    \boxminus_0&=&e^2\mathfrak{Re}\int\frac{d^4q_E}{(2\pi)^4}\frac{M_W^2}{M_W^2+Q^2}\frac{1}{(Q^2)^2}\frac{Q^2-\nu_E^2}{\nu_E}\frac{T_3(i\nu_E,Q^2)}{M M_F}~,\nonumber\\
    \boxminus_1&=&-\frac{8}{3}e^2\mathfrak{Re}\int\frac{d^4q_E}{(2\pi)^4}\frac{Q^2-\nu_E^2}{(Q^2)^3}\frac{iT_3(i\nu_E,Q^2)}{M M_F}~,\label{eq:boxminus}
\end{eqnarray}
where $q_E=(\vec{q},\nu_E)$ is the Euclidean loop momentum, and $Q^2=\q^2+\nu_E^2$. 
It is clear that $\boxminus_0$ and $\boxminus_1$ probe the odd and even component of $T_3$ under $\nu\rightarrow-\nu$, respectively:
\begin{equation}
    T_{3,\pm}(\nu,Q^2)\equiv\frac{1}{2}(T_3(\nu,Q^2)\pm T_3(-\nu,Q^2))~.
\end{equation}

Eq.\eqref{eq:boxminus} may serve as the starting point for nuclear theory calculations~\cite{NCSMC10}, but since $T_3$ involves a time-ordered product of two currents (or the nuclear Green's function in the momentum space), it may not be the simplest quantity to work on. A convenient alternative is to make use of the dispersion relation for $T_{3,\pm}$ with respect to the variable $\nu$:
\begin{eqnarray}
iT_{3,-}(\nu,Q^2)=4\nu\int_{\nu_0}^\infty d\nu'\frac{F_{3,-}(\nu',Q^2)}{\nu^{\prime 2}-\nu^2}\,,\quad
iT_{3,+}(\nu,Q^2)=4\nu^2\int_{\nu_0}^\infty d\nu'\frac{F_{3,+}(\nu',Q^2)}{\nu'(\nu^{\prime 2}-\nu^2)}~,
\end{eqnarray}
with $\nu_0>0$ the inelastic threshold. Here, we introduced the structure (response) functions, 
\begin{align}
F_{3,\pm}(\nu,Q^2)&=-\frac{iM\nu}{2\q}\sum_X\delta(E_X-M-\nu)\nonumber\\
&\times\left\{\langle \phi_f|J_\text{em}^x(\vec{q})|X\rangle \langle X|(J_W^{\dagger y})_A|\phi_i\rangle\mp \langle \phi_f|(J_W^{\dagger y}(-\vec{q})_A|X\rangle\langle X|J_\text{em}^x(\vec{q})|\phi_i\rangle\right\}\,.
\label{eq:F3currents}
\end{align}
Nuclear response functions are standard objects of nuclear ab-initio calculations, e.g. in studies of lepton-nucleus scattering~\cite{Shen:2012xz,Lovato:2014eva,Lovato:2017cux,Rocco:2020jlx,Lovato:2020kba,Sobczyk:2021dwm,Sobczyk:2023mey}.
Plugging them back to Eq.\eqref{eq:boxminus} gives the dispersive representation~\cite{Seng:2018yzq,Gorchtein:2018fxl}:
\begin{eqnarray}
    \boxminus_0&=&\frac{\alpha}{\pi}\int_0^\infty dQ^2\frac{M_W^2}{M_W^2+Q^2}\int_{\nu_0}^\infty\frac{d\nu}{\nu}\frac{\nu+2\sqrt{\nu^2+Q^2}}{(\nu+\sqrt{\nu^2+Q^2})^2}\frac{F_{3,-}(\nu,Q^2)}{M M_F}\nonumber\\
    \boxminus_1&=&\frac{2\alpha}{3\pi}\int_0^\infty dQ^2\int_{\nu_0}^\infty\frac{d\nu}{\nu}\frac{\nu+3\sqrt{\nu^2+Q^2}}{(\nu+\sqrt{\nu^2+Q^2})^3}\frac{F_{3,+}(\nu,Q^2)}{M M_F}~.\label{eq:boxminusDR}
\end{eqnarray}
In the second row, we suppressed the remnant of the $W$-boson propagator because the respective $\nu$-integral is insensitive to large loop momenta. 
For practical calculations, it is convenient to expand current operators into multipole series, which results in a multipole expansion of $T_{3,\pm}$ and $F_{3,\pm}$; explicit expressions can be found in Eqs.76, 77 of Ref.\cite{Seng:2022cnq}.


\subsection{Subtracting the single-nucleon contribution in $\boxminus_0$}


Among the different terms in $\Box_{\gamma W}(E_e)$, only $\boxminus_0$ occurs simultaneously in both nucleon and nuclei.
In fact, it is the only piece that is sensitive to all loop momenta $0<Q^2<\infty$; for $\delta_\text{NS}$ we need the difference between the nuclear and nucleon version of this quantity.  

The current status of $\boxminus_0^n\approx \Box_{\gamma W}^n$ is summarized in Sec.\ref{sec:Boxn}. A direct computation of $\boxminus_0^\text{nucl}$
in a nuclear many-body approach, however,  is hardly possible, not least because all existing computational methods in nuclear theory only operate with \textit{nucleonic} degrees of freedom, and thus are only applicable to a limited region of low loop momenta. We label the outcome of a nuclear many-body calculation 
as $(\boxminus_0^\text{nucl})_\text{MB}$ to distinguish it from the full $\boxminus_0^\text{nucl}$. Then, the value of $\delta_\text{NS}$ should be inferred from the knowledge of $(\boxminus_0^\text{nucl})_\text{MB}$ and $\boxminus_0^n$. 
\begin{figure}
    \centering
\includegraphics[width=0.75\columnwidth]{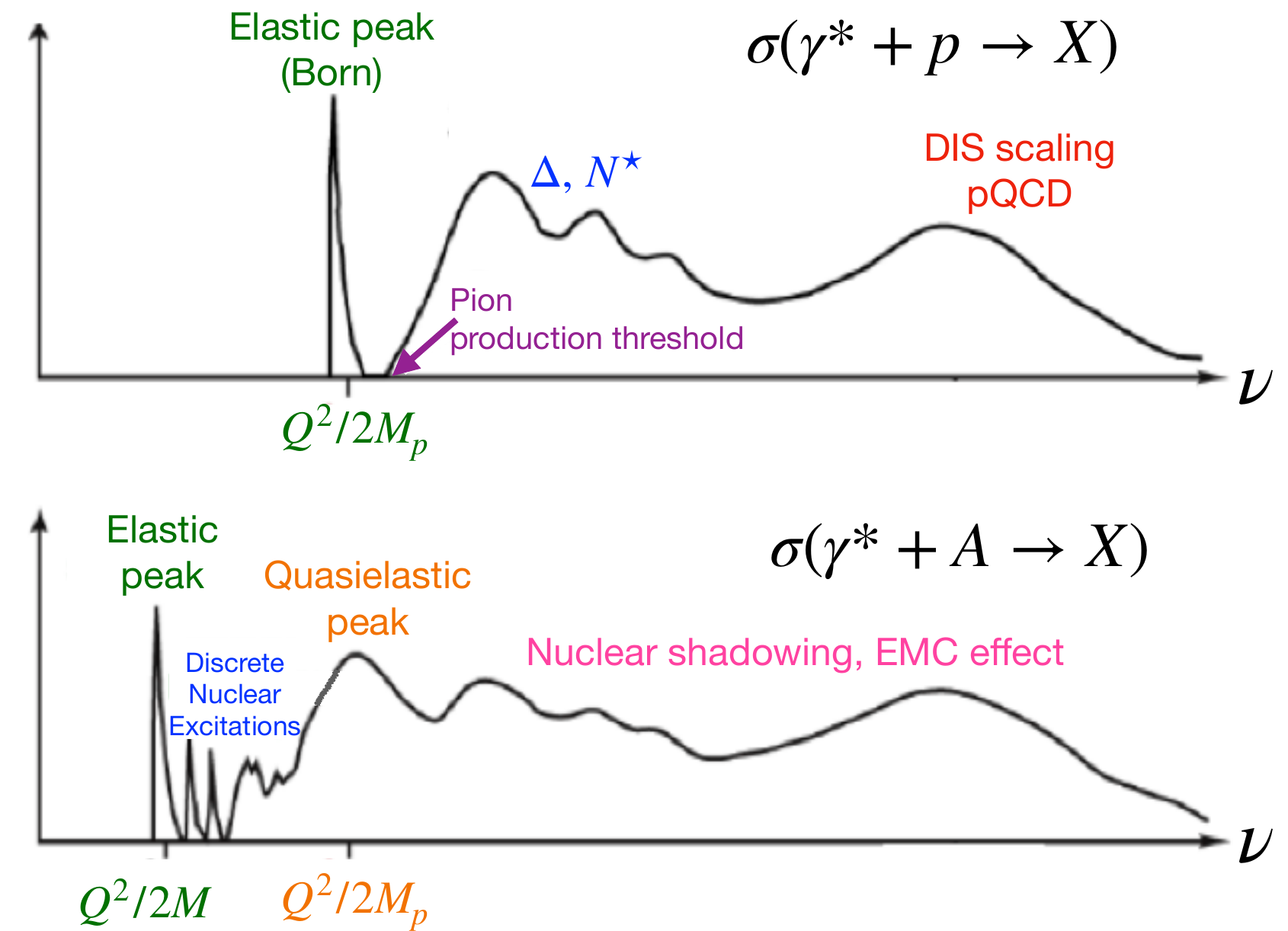}
    \caption{
    A schematic representation of the nucleon (above) and nuclear (below) virtual photoabsorption.}
    \label{fig:absorption}
\end{figure}

The dispersive approach is well suited 
to answer this question. 
From the dispersion representation of $\boxminus_0$ in Eq.\eqref{eq:boxminusDR} it is clear that the difference between the nuclear and nucleon case is encoded in the structure function: various contributions to it may differ in the {\it position} and {\it strength}, depending on whether a free nucleon or a nucleus is considered.
Fig.\ref{fig:absorption} shows a schematic comparison of the inclusive virtual photoabsorption spectrum on a free nucleon and a nucleus for a fixed value of $Q^2$ and as a function of the virtual photon energy $\omega$ (equivalent to$\nu$).
Going 
from low to high energy, the first contribution we encounter in the case of the free nucleon is that due to the nucleon ground state (elastic, or Born contribution), and the lowest inelastic state is separated from it by the pion mass. All features of the absorption spectrum on a free nucleon above the pion threshold are also present for nuclei. At the same time, the absorption spectra below pion threshold are definitely different. 
It has been common to use the pion production threshold as the watershed between $\Delta_R^V$ and $\delta_\text{NS}$: All heavier states' contributions are included in the former and assumed to be nucleus-independent, while nuclear-specific contributions only reside at low energies, and only nucleons are considered dynamical. 
Under these assumptions, the subtraction of the single-nucleon box is realized as
\begin{equation}
\boxminus_0^\text{nucl}-\boxminus_0^n\approx (\boxminus_0^\text{nucl})_\text{MB}-(\boxminus_0^n)_\text{Born}~. \label{eq:BornSubt}
\end{equation}
This definition assumes that all non-nucleonic contributions to $\boxminus_0$ ($N\pi$, resonance, Regge, DIS, and so on) receive no modification in the nuclear medium. The extreme asymptotic part is fixed by (the $\gamma W$-interference analog of) the Gross-Llewellyn Smith (GLS) sum rule~\cite{Gross:1969jf}. Although the nuclear EMC effect modifies the $x$-dependence of DIS structure functions in a nucleus at large $Q^2$ compared to free nucleon~\cite{EuropeanMuon:1983wih,Gomez:1993ri,CLAS:2019vsb,Seely:2009gt}, the GLS sum rule value remains unchanged~\cite{Leung:1992yx,Kim:1998kia}. All sub-asymptotic contributions undergo a nuclear shadowing, and we refer the reader to a recent review~\cite{Kopeliovich:2012kw}. We stress that the nuclear modification effect on the inelastic contributions to the $\gamma W$-box correction to nuclear beta decay rates has never been either considered or articulated in the literature.


\subsection{\label{sec:shortcoming}Estimates of $\delta_\text{NS}$ in the literature}

\begin{figure}
    \centering
    \includegraphics[width=0.5\columnwidth]{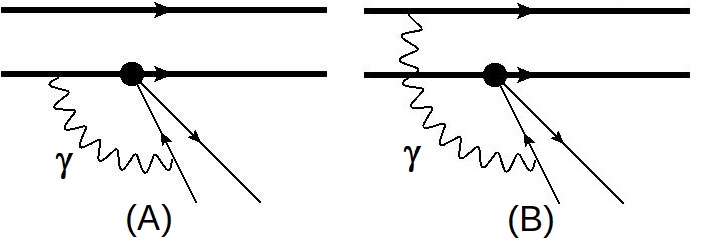}
    \caption{
    Representative diagrams for the traditional splitting of $\delta_\text{NS}$ into $\delta_\text{NS}^A$ and $\delta_\text{NS}^B$. 
    }
    \label{fig:deltaNS}
\end{figure}
Historically, only the first, energy-independent term of Eq.\eqref{eq:deltaNSexpand} was considered, with the formal definition in Eq.\eqref{eq:BornSubt}. If operating only with one-body electroweak nucleon currents embedded in a nucleus, the only two possibilities are represented by the schematic 
diagrams in Fig.\ref{fig:deltaNS}. The first identified nuclear effect, absent in the single-nucleon RC, stems from the diagram (B)~\cite{Jaus:1989dh} where the weak and EM vertices act on two different nucleons. Its contribution was estimated in the nuclear shell model~\cite{Barker:1991tw,Towner:1992xm}. Later, Towner argued~\cite{Towner:1994mw} that since the single-nucleon Gamow-Teller (GT) and magnetic coupling constants were observed to be reduced  (quenched) in nuclear medium~\cite{Brown:1983zzc,Brown:1985zz,Brown:1987obh,Towner:1987zz},
the diagram (A), where the weak and EM vertices act on the same nucleon, would be modified. Its contribution was estimated by applying the quenching factors $q_M$ and $q_{GT}$ to the magnetic and axial vertices, respectively, in the Born term in $\Box_{\gamma W}^n$. This results in $\delta_\text{NS}^A\equiv2(q_Mq_{GT}-1)\Box_{\gamma W}^{n,\,\text{Born}}$ being negative-definite, since quenching factors are smaller than 1. The quenching effect was later also introduced to the two-nucleon diagram~\cite{Towner:2002rg,Towner:2007np}. Unlike $\delta_\text{NS}^A$, the sign of $\delta_\text{NS}^B$ was found to alternate. 
The two corrections formed the standard know-how adopted until 2018. To this date, the estimates of Hardy and Towner~\cite{Hardy:2014qxa} are the only ones to address both $\delta_\text{NS}^A$ and $\delta_\text{NS}^B$. These shell-model calculations did not account for any specifically nuclear feature of the absorption spectrum (e.g., discrete states, giant resonances, and such). The loop integral was performed with nucleon magnetic and axial form factors, just like the Born contribution in the free-nucleon case, which suggests a free-nucleon propagating between the vertices, without interaction with the spectator. Because the quenching factors explicitly refer to transitions between {\it nuclear states}, not {\it nucleus to continuum}, their simultaneous use with the nucleon form factors in the weak axial and magnetic vertices in the approach of Refs.\cite{Jaus:1989dh,Towner:1992xm,Towner:1994mw} is inconsistent. Moreover, the modern understanding of the quenching phenomenon is its being an artifact of the nuclear shell model with only one-body currents: no quenching is required if two- and many-body effects are included~\cite{Gysbers:2019uyb,Coraggio:2018tuo}.


The first dispersion relation-based calculation of $\delta_\text{NS}$ \cite{Seng:2018qru} could only assess $\delta_\text{NS}^A$. It identified it with the replacement of the free-nucleon Born contribution with a bound-nucleon one. This replacement is illustrated in Fig.\ref{fig:absorption} where the sharp elastic peak on the free nucleon is replaced by a broad quasielastic peak on the nucleus. A simple estimate in the free Fermi gas model indicated that the quenching factor-based calculation of Ref.\cite{Towner:1994mw} underestimated $\delta_\text{NS}^A$ by about a factor 2, a shift about 3 times the total uncertainty in the $\mathcal{F}t$ analysis. A subsequent application of this model to the energy-dependent term $\boxminus_1$ \cite{Gorchtein:2018fxl} questioned the neglect of the energy dependence in $\delta_\text{NS}$. Surprisingly, it found that an inclusion of this energy dependence largely cancelled the shift found in Ref. \cite{Seng:2018qru}, leaving the central value unchanged. However, because of this cancellation, and since both estimates were based on an unsophisticated free Fermi gas model, a conservative 100\% uncertainty was assigned to the entire effect, triplicating the uncertainty of $\mathcal{F}t$ and the nuclear-structure uncertainty of $V_{ud}$ extracted from the superallowed decays. We  refer the reader to Table XI of Ref.\cite{Hardy:2020qwl} for the current status of $\delta_\text{NS}$, as summarized in this subsection.

\subsection{Singular terms}

Finally, we discuss a contribution which is unique to some superallowed decays, the residue contribution from $T_3$~\cite{Gennari}.
For five superallowed decays, $^{10}\text{C}\rightarrow {}^{10}\text{B}$, $^{14}\text{O}\rightarrow {}^{14}\text{N}$, $^{18}\text{Ne}\rightarrow {}^{18}\text{F}$, $^{22}\text{Mg}\rightarrow {}^{22}\text{Na}$ and $^{30}\text{S}\rightarrow {}^{30}\text{P}$, the $T_z=0$, $J^P=0^+$ state is not a ground state, and low-lying states with $T=0$ and $J^P=1^+$ are present \cite{ENDF-VIIshort}. 
If the initial and final $0^+$ states can decay to such a low-lying state $k$ with $M_k<M$,
it will lead to a pole of $T_3$ in the third quadrant at  $\nu_k=\sqrt{M_k^2+\q^2}-M-i\epsilon$, with 
the residue:
\begin{equation}
\mathfrak{Res}T_3^\text{nucl}(\nu_k,\q)=\frac{2M\nu_k}{\q}\sum_{s_k}\langle\phi_f|J_\text{em}^x(\vec{q})|k,\vec{q},s_k\rangle\langle k,\vec{q},s_k|J_{W5}^{\dagger y}(-\vec{q})|\phi_i\rangle~.
\end{equation}
Upon performing the Wick rotation, the $\gamma W$-box will receive the residue contribution,
\begin{eqnarray}
(\Box_{\gamma W}^\text{nucl}(E_e))_{\text{res},T_3}&=&\frac{e^2}{M}\sum_k\mathfrak{Re}\int_0^{\q_\text{max}}\frac{d\q \q^2}{(2\pi)^2}\frac{1}{\q^2-\nu_k^2}\frac{i\mathfrak{Res}T_3^\text{nucl}(\nu_k,\q)}{M_F}\nonumber\\
&&\times \left\{\frac{2|\vec{p}_e|^2\q^2+\nu_k E_e\mathcal{A}}{4\nu_k |\vec{p}_e|^3\q}\ln\left|\frac{\mathcal{A}+2|\vec{p}_e|\q}{\mathcal{A}-2|\vec{p}_e|\q}\right|-\frac{E_e}{|\vec{p}_e|^2}\right\}~,\label{eq:residue}
\end{eqnarray}
with $\mathcal{A}=\nu_k^2-2E_e\nu_k-\q^2$ and $\q_\text{max}=\sqrt{M^2-M_k^2}$.

\begin{figure}
    \centering
    \includegraphics[width=0.75\columnwidth]{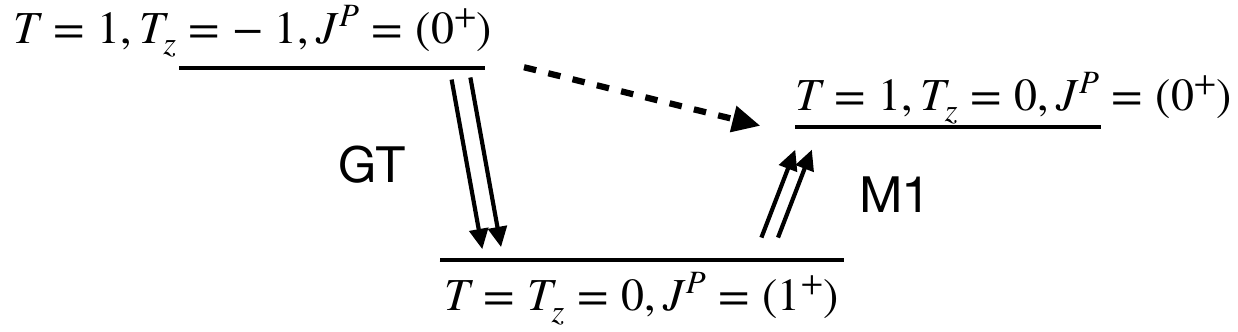}
    \caption{
    The presence of a low-lying $1^+$ state in the daughter nucleus that leads to the residue contribution in $\delta_\text{NS}$. The superallowed transition $i\to f$ is indicated by the dashed arrow. A two-step process $i\to k\to f$ by means of subsequent GT and M1 transitions is indicated by solid double arrows.}
    \label{fig:residue}
\end{figure}

Physically, it corresponds to a two-step transition through a ground or another excited state of the daughter nucleus that lies below the $0^+$ one, as shown in Fig.\ref{fig:residue}. 
Unlike the other terms in Eq.\eqref{eq:deltaNSexpand}, this contribution contains negative powers of $E_e$. 
Although this singular behavior is integrable, 
it can lead to a non-trivial distortion of the beta spectrum and is clearly of great importance. 
%
For a simple estimate, the nuclear matrix elements can be inferred from experimental half-lives of the M1 transition $\phi_f(0^+)\rightarrow k(1^+)$ and the GT transition $\phi_i(0^+)\rightarrow k(1^+)$,
%
\begin{equation}
|M_\text{M1}|^2=\frac{3}{4\alpha}\frac{\ln2}{t_\text{M1}E_\gamma^3}~,~|M_\text{GT}|^2=\frac{2\pi^3\ln 2}{G_F^2V_{ud}^2m_e^5(ft)_\text{GT}}~.
\end{equation}
Above, $(ft)_\text{GT}$ is the $ft$-value of the GT-transition,  $t_\text{M1}$ is the half-life of the M1 transition, and $E_\gamma$ its energy. This fixes the residue up to an overall sign:
\begin{equation}
\frac{i\mathfrak{Res}T_3^\text{nucl}(\nu,\q)}{M_F}=\pm\sqrt{\frac{\pi^3M^2\nu_k^2(\ln 2)^2}{3G_F^2V_{ud}^2\alpha m_e^5E_\gamma^3(ft)_\text{GT}t_\text{M1}}}\left(1+\mathcal{O}(\q)\right)\label{eq:resfromdata}
\end{equation}
where $\mathcal{O}(\q)$ denotes extra $\q$-dependence in the nuclear matrix elements between $0<\q<\q_\text{max}$ not captured by the decay processes due to the smallness of the phase space. 
\begin{table*}[h]
\caption{\label{tab:GTM1}A collection of the relevant $ft$-values for the GT-transitions and M1 transition half-lives for five superallowed decay channels. Data from Ref.~\cite{ENDF-VIIshort}. The rightmost column summarizes estimates of the respective residue contribution. }
	\begin{centering}
\begin{tabular}{c|c|c|c}
\hline 
 & GT, $\log_{10}ft$ (s) & M1, $t_{1/2}$
 &$2|\langle (\Box_{\gamma W}^\text{nucl}(E_e))_{\text{res},T_3}\rangle|$
 \tabularnewline
\hline 
\hline 
$^{10}$C$\rightarrow$$^{10}$B & 3.0426(7) & 4.9(21) fs & $4.2^{+1.4}_{-0.8}\times10^{-3}$\tabularnewline
\hline 
$^{14}$O$\rightarrow$$^{14}$N & 7.279(8) & 68(3) fs & $4.6^{+0.2}_{-0.0}\times10^{-6}$\tabularnewline
\hline 
$^{18}$Ne$\rightarrow$$^{18}$F & 3.091(4) & 1.77(31) fs & $9.0^{+1.0}_{-0.6}\times 10^{-3}$\tabularnewline
\hline 
$^{22}$Mg$\rightarrow$$^{22}$Na & 3.64 & 19.6(7) ps & $7.0^{+0.2}_{-0.2}\times10^{-4}$\tabularnewline
\hline 
$^{30}$S$\rightarrow$$^{30}$P & 4.322(11) & 96(10) fs & $5.8^{+0.4}_{-0.2}\times10^{-4}$\tabularnewline
\hline 
\end{tabular}
\par\end{centering}
\end{table*}

In Table~\ref{tab:GTM1} we summarize the current experimental data on the GT and M1 transition rates of interest, and combine them to obtain the estimate for the residue contribution to $\delta_\text{NS}$. Apart from the $^{14}\text{O}\to {}^{14}\text{N}$ transition, the effect of the residue contribution on the other four ones is remarkably large. This can be understood by noting that while a typical superallowed transition half-life is of the order of a few seconds, the M1 half-lives are a few femtoseconds, so even though the two-step process is generally suppressed with respect to the direct one, the much higher electromagentic transition rate makes this contribution sizable. The uncertainties of the estimates of $2|\langle (\Box_{\gamma W}^\text{nucl}(E_e))_{\text{res},T_3}\rangle|$ in Table \ref{tab:GTM1} stem entirely from those in the M1 and GT rates and disregard all other uncertainties which can be significant. 
While the GT rates are very precisely measured, the M1 ones are more uncertain, and are worth determining with a better precision. The general structure and details of the residue contribution appear in this review for the first time. The final word on its size and sign should be said by ab-initio nuclear calculations.


\section{Summary: superallowed decays at the crossroads}

The current status of $V_{ud}$ extraction from superallowed nuclear beta decays stems from the latest ``Critical survey'' by Hardy and Towner~\cite{Hardy:2020qwl}:
\begin{equation}
|V_{ud}|_{0^+}=0.97361(5)_\text{exp}(6)_{\delta_R^\prime}(4)_{\delta_\text{C}}(28)_{\delta_\text{NS}}(10)_\text{RC}[31]_\text{total}~.\label{eq:Vud0+}
\end{equation}
Combined with $|V_{us}|=0.2243(8)$ from kaon decays channels~\cite{ParticleDataGroup:2022pth}, it returns $\Delta_u^{0+} \equiv|V_{ud}|^2_{0+}+|V_{us}|^2-1=-0.00166(69)$ which indicates a $2.4\sigma$ unitarity-deficit. This is to be compared to neutron decay~\cite{Gorchtein:2023srs}:  
$\Delta_u^{\text{n,\,PDG-av}}=-0.00037(174)$
which shows no such deficit, a mild disagreement which is to be understood.

The possibility to jointly analyze many decay channels makes superallowed decays unique: no experimental input other than the superallowed decays themselves is required to set stringent constraints on BSM-induced scalar 4-Fermi operators \cite{Lee:1956qn,Jackson:1957zz,Falkowski:2020pma}. The latter manifest themselves as the Fierz interference term $b_F$ which would reintroduce the $Z$-dependence in the $\mathcal{F}t$-values,  
\begin{equation}
    \Big[\mathcal{F}t\Big]^\text{BSM}
    \sim \Big[\mathcal{F}t\Big]^0
    \left[1+b_F\langle\frac{m_e}{E_e}\rangle\right]^{-1},\label{eq:Fierz}
\end{equation}
since $\langle E_e\rangle\sim Q_\text{EC}$ and the $Q_\text{EC}$ grows with the growing $Z$. The internal consistency of the superallowed decay data base with the constant-$\mathcal{F}t$ fit is thus per se a sensitive test of SM and beyond. The latest Hardy-Towner survey quoted $|b_F|\leqslant 0.0033$ at the 90\% confidence level. Because in the formalism of Hardy and Towner $\delta_\text{C}$ is the largest nuclear structure-dependent correction with the strongest $Z$-dependence, the observed model dependence of this correction~\cite{Hardy:2014qxa,Xayavong:2017kim} poses a problem for the entire analysis. Hardy and Towner then proposed to use the constancy of $\mathcal{F}t$ values as an additional requirement on the nuclear models, and only after the constant fit is preformed assess the limits on the Fierz term. This approach is inconsistent since the constant $\mathcal{F}t$'s is the mathematical expression of the absence of scalar BSM. 
On the other hand, while a simultaneous fit to all $\mathcal{F}t$-values with $V_{ud}$ and $b_F$ as free parameters seems a more coherent choice, it would result in a large nuclear model dependence of both extracted parameters, and is hardly an optimal solution. 


To summarize this review: It was found that some aspects of the theoretical formalism that served as the basis for $V_{ud}$ extraction since the dawn of the SM, do not match the current $10^{-4}$ precision requirements for the precision test of SM. In the past few years a significant effort was put in reassessing this formalism without resorting to the historically used approximations, and to addressing the nuclear and hadronic uncertainties more rigorously. The novel formalism based on dispersion theory is an appropriate tool for this reassessment, as it allows to accommodate inputs from experimental data, perturbative and lattice QCD, Regge phenomenology, and effective field theory in the nucleon and nuclear sector. 
We reviewed all relevant ingredients needed to convert experimental measurements to a precise value of $V_{ud}$ and limits on BSM. 
With the new model-independent and complete formalism for $\delta_\text{NS}$ presented here, everything is ready for applications of modern nuclear-theoretical methods to computations of $\delta_\text{NS}$ with a robust uncertainty estimate. We showed how data-driven approach can help taming nuclear-structure uncertainties. 
Future high-precision measurements of nuclear radii will help significantly reduce nuclear uncertainties in the subleading corrections to the Fermi function and the ISB correction $\delta_\text{C}$. The latter, additionally, should be newly explored with the best theoretical tools we have at hand. 

Because some of these uncertainties have been erroneously neglected in the past, the new approach has momentarily led to their increase. 
However, armed with a novel, more advanced formalism, we envision a significant improvement in $V_{ud}$ extraction and scalar BSM limits in the near future. 

\section*{ACKNOWLEDGMENTS}

The authors acknowledge helpful discussions with S. Bacca, F. Bonaiti, M. Drissi, M. Gennari, G. Hagen, M. Hoferichter, V. Lubicz, P. Navratil, B. Ohayon, R. Pohl, J. Sobczyk and F. Wauters.
The work of M.G. is supported in part by EU Horizon 2020 research and innovation programme, STRONG-2020 project
under grant agreement No 824093, and by the Deutsche Forschungsgemeinschaft (DFG) under the grant agreement GO 2604/3-1. The work of C.-Y.S. is supported in
part by the U.S. Department of Energy (DOE), Office of Science, Office of Nuclear Physics, under the FRIB Theory Alliance award DE-SC0013617, by the DOE grant DE-FG02-97ER41014, and by the DOE Topical Collaboration ``Nuclear Theory for New Physics'', award No.
DE-SC0023663.

\bibliographystyle{ar-style5.bst}

\bibliography{NuclearBeta.bbl}

\begin{thebibliography}{150}
\bibitem{NCSMC10}
Gennari M, et~al. To be published.
\end{thebibliography}




\end{document}